\documentclass[useAMS,referee]{mn2e}
\usepackage{psfig,epsfig}

\begin{document}
\def\lsun{{\rm L_{\odot}}}
\def\msun{{\rm M_{\odot}}}
\def\rsun{{\rm R_{\odot}}}
\def\p{$e^\pm \;$}
\def\Pacz{Paczy\'nski~}
\def\go{
\mathrel{\raise.3ex\hbox{$>$}\mkern-14mu\lower0.6ex\hbox{$\sim$}}
}
\def\lo{
\mathrel{\raise.3ex\hbox{$<$}\mkern-14mu\lower0.6ex\hbox{$\sim$}}
}
\def\simeq{
\mathrel{\raise.3ex\hbox{$\sim$}\mkern-14mu\lower0.4ex\hbox{$-$}}
}

\input epsf.sty

\title{High Resolution Calculations of Merging Neutron Stars II: 
Neutrino Emission}
\author[Rosswog \& Liebend\"orfer]
       {S. Rosswog$^{1}$ and M. Liebend\"orfer$^{2}$
\\${\bf 1.}$ Department of Physics and Astronomy,University of Leicester, 
        LE1 7RH, Leicester, UK
\\${\bf 2.}$ CITA, University of Toronto, Toronto, Ontario M5S 3H8, CA}
\maketitle

\begin{abstract}

The remnant resulting from the merger of two neutron stars produces 
neutrinos in copious amounts. In this paper we present the neutrino emission 
results obtained via Newtonian, high-resolution simulations of the coalescence
event. These simulations use three-dimensional smoothed particle hydrodynamics
together with a nuclear, temperature dependent equation of state and 
a multi-flavour neutrino leakage scheme. We present the details of our
scheme, discuss the neutrino emission results from a neutron star coalescence
and compare them to the core-collapse supernova case where neutrino emission 
has been studied for several decades.
The average neutrino energies are similar to those in the supernova case, 
but contrary to the latter, the luminosities are dominated by electron-type 
antineutrinos which are produced in the hot, neutron-rich, thick disk of 
the merger remnant. The cooler parts of this disk contain substantial 
fractions of heavy nuclei, which, however,  do not influence the overall 
neutrino emission results significantly.
Our total neutrino luminosities from the merger event are considerably 
lower than those found in previous investigations. This has serious consequences
for the ability to produce a gamma-ray burst via neutrino annihilation.
The neutrinos are emitted preferentially along the initial binary rotation 
axis, an event seen ``pole-on'' would appear much brighter in neutrinos than 
a similar event seen ``edge-on''.

\end{abstract}


\section{Introduction}
Binary pulsars such as the famous  PSR 1913+16  are fascinating laboratories
for extreme physics. Soon after its discovery it was realized that the orbit
of PSR 1913+16 is decaying due to energy constantly leaking out of the 
system in the form of gravitational waves (Taylor 1994) and therefore 
making the final coalescence an inescapable consequence. 
This merger event holds promises for areas as diverse as gamma-ray bursts
(\Pacz 1986, Eichler et al. 1989, Narayan et al. 1992), ground-based 
gravitational wave
detection (Abramovici et al. 1992, Kuroda et al. 1997, Bradaschia et al. 1990,
Danzmann 1997) and the formation of rapid neutron capture elements 
(Lattimer \& Schramm 1974, Lattimer \& Schramm 1976, Symbalisty \& Schramm 1982, 
Eichler et al. 1989, Rosswog et al. 1999, Freiburghaus et al. 1999).\\
The involved physics of the event is in almost every aspect ``exotic'':
the neutron star fluid moves in and determines the dynamical,
curved space-time; in the centers of the stars and the resulting merger 
remnant the baryon densities reach multiples of the nuclear saturation density,
$\rho_s = 2.5 \cdot 10^{14}$ gcm$^{-3}$;
nuclear reactions proceed via extremely neutron-rich and short-lived isotopes;
and the initial neutron star magnetic fields are expected to be amplified
during the merger to a strength, $B\sim 10^{17}$ G, so that their feed-back 
on the fluid flow becomes dynamically important 
(Thompson \& Duncan 1993, Thompson 1994, Kluzniak \& Ruderman 1998, 
Rosswog \& Davies 2002).\\
Neutron star mergers represent
a severe challenge for computer simulations. The event is genuinely 
multidimensional and unlike, for example, with core-collapse supernovae (SNe), 
there are no basic open questions that
could be first addressed in restricted dimensions, as e.g. the robustness
of the delayed neutrino-driven supernova mechanism. Nevertheless,
the two events have several common aspects: compact objects are formed
in the center of the event and huge amounts of gravitational binding energy,
of order $10^{53}$ erg, are released in form of neutrinos making neutrino 
physics a key ingredient of both scenarios. The material in the 
innermost layers of both configurations 
is very dense, neutron rich, and neutrino opaque. Most neutrinos are
radiated from a hot and thick accretion disk in the neutron star merger
case, and from a shock heated mantle in the standard supernova scenario.
The neutrino emission and absorption are the key features in the picture
of a neutrino-driven supernova, which has been sketched in the sixties
(Colgate \& White 1966, Arnett 1967, Schwartz 1967), then refined in
the mid-eighties (Wilson 1971, Arnett 1977, VanRiper \& Lattimer 1981,
Bowers \& Wilson 1982, Wilson 1985, Bethe \& Wilson 1985, Bruenn 1985),
and still continues to be controversially discussed and improved by
many researchers. One result of this effort is the emergence of 
sophisticated neutrino transport schemes (although currently restricted
to low spatial dimensions) to address the viability of the neutrino-driven
supernova model. Starting with leakage schemes that considered only
neutrino emission (VanRiper \& Lattimer 1981, Baron et al. 1985),
multi-group flux limited diffusion approximations 
(Arnett 1977, Bowers \& Wilson 1982, Bruenn 1985, Myra et al. 1987,
Bruenn et al. 2001)
have been developed that take the energy spectra and a truncated expansion
in the propagation direction between emission and absorption into
account. While multidimensional simulations relying on transport approximations
with externally imposed neutrino fluxes or spectra (Herant et al. 1994,
Burrows et al. 1995, Janka \& M\"uller 1996, Mezzacappa et al. 1998,
Fryer \& Warren 2002)
throwed a bridge between simulation and observation, the traditional
investigations in spherically symmetric geometry proceeded to solutions
of the complete Boltzmann transport equation in stellar core collapse
(Mezzacappa \& Bruenn 1993) and postbounce evolution (Rampp \& Janka 2000,
Mezzacappa et al. 2001, Burrows \& Thompson 2002), including
full general relativity (Liebend\"orfer et al. 2001). The coalescence
of neutron stars occurs on a much shorter
time scale, of order milliseconds, compared to the shock revival in
a supernova, which is believed to take several tenths of a second.
Although weak interactions provide an important mechanism for the cooling
of the disk that is opaque to all forms of electro-magnetic radiation, they
do not allow for dramatic changes in the temperature and electron
fraction, at least not on time scales accessible to current numerical
simulations. Multidimensional kinematics seems to remain the dominant 
ingredient of neutron star mergers.\\
Due to the complexity of the event simulations are still divided into
two classes: either focussing on the strong-field gravity aspect 
(Oohara \& Nakamura 1997, Ayal et al. 2001, Faber \& Rasio 2000, Faber et al.
2001, Wilson et al. 1996 , Baumgarte et al. 1997,
Oechslin et al. 2002, Shibata 1999, Shibata \& Uryu 2000, 
Shibata \& Uryu 2002) thereby sacrificing
possibly important microphysics or exploring microphysics but using
essentially Newtonian gravity 
(Ruffert et al. 1996, Ruffert et al. 1997, Ruffert \& Janka 2001, Rosswog et
al. 1999, Rosswog et al. 2000, Rosswog \& Davies 2002, Rosswog \& Ramirez-Ruiz
2002). Neutrino physics has, to our knowledge, so far only been 
included in the simulations of Ruffert et al. (see Ruffert \& Janka 2001
and references therein) and in  Rosswog \& Davies (2002). \\
In this paper, we detail on our neutrino leakage scheme that has been 
used in our high-resolution, three-dimensional simulations 
 of merging neutron stars and report on the corresponding neutrino
emission results. Our leakage
scheme is meant to join the current state-of-the-art for this specific,
three-dimensional application where simplicity and numerical efficiency 
are valuable assets. It is not supposed to compare with much more 
elaborate (but low-dimensional) transport schemes necessary for 
quantitative statements about possibly neutrino-driven
supernovae.  Knowing how important the stiff energy
dependence of the weak interactions is in the supernova, we design
the leakage scheme to avoid the usage of mean energies for the determination
of neutrino source functions or opacities. We determine for each neutrino
energy separately a production rate and a diffusion time scale. The
latter depends on a non-local estimate for the optical depth from which
we extract the explicit energy dependence. The rates for the production
of new neutrinos and the diffusion of neutrinos from local equilibrium
are then analytically integrated over energy. The smoothed minimum
of production and diffusion rates is used as leakage source in
the hydrodynamics equations. We apply this procedure separately for
the lepton number and energy transfer.\\
In Section 2 we will summarize previous results, in Section 3 we report
on the neutrino emission results from our merger simulations. The 
summary and a discussion of the results is provided in Section 4 and 
the details of the neutrino treatment are given in the Appendix.

\section{Basic model features and previous results}

%
%
We have  performed a set of high-resolution simulations of the last 
inspiral stages and the final coalescence of a double neutron star system.
Large parts of the model and the hydrodynamic evolution have been described
in detail in Rosswog \& Davies (2002), hereafter referred to as paper I.
The numerical runs analyzed in this paper are, apart from additional test
runs, the same as those described in paper I. Here we focus on 
the parts of the model and the results that are related to the emission 
of neutrinos.

Keeping in mind its decisive role for the (thermo-)dynamical evolution
of the merger event (see e.g. Rosswog et al. 1999, Rosswog et al. 2000)
we use  an equation of state (EOS) for hot and dense nuclear matter. 
Our equation of state is based 
on the tables provided by Shen et al. (1998a, 1998b). We have added the 
lepton and photon contributions, and extended it smoothly to the 
low-density regime with a gas consisting of neutrons, alpha particles, 
electrons, positrons and photons. For details concerning the EOS we refer 
to paper I. 
The Newtonian self-gravity of the fluid is calculated efficiently via a binary 
tree (Benz et al. 1990). The back-reaction forces that emerge from 
the emission of gravitational waves are added in the point-mass limit of 
the quadrupole approximation.

To solve the equations of hydrodynamics for the neutron star fluid 
we have applied the smoothed particle hydrodynamics method (SPH; e.g. 
Benz (1990) or Monaghan (1992)). 
It is a widespread misconception that
SPH is viscous ``by nature'' and thus necessarily introduces artefacts 
in simulations
of low-viscosity flow. First, the degree of viscosity present in SPH is, 
as in every numerical scheme, a function of the numerical resolution.
The components of the SPH artificial viscosity tensor scale to
leading order proportional to the smoothing length $h$, which tends to zero 
with increasing resolution. 
The standard form of the SPH artificial viscosity tensor 
(e.g. Monaghan (1992)) is known to introduce spurious
forces in pure shear flows. We have applied a switch suggested by Balsara
(1995) which suppresses these forces in case of pure shear and reproduces 
the original form in case of shocks. A further improvement concerns the
artificial viscosity parameters, usually called $\alpha$ and $\beta$: they
are made time dependent (as suggested in Morris and Monaghan 1997) and 
an additional differential equation
is solved to determine their values. In the absence of shocks these values
are negligible, if a shock is detected the parameters rise to their standard 
values. This artificial viscosity treatment is described and tested in 
detail in Rosswog et al. (2000).\\
To quantify the amount of viscosity in our current simulations we have 
estimated the effective $\alpha$-viscosity present in the disk of the 
merger remnant. The effective $\alpha$-viscosity is
 $\alpha_{SS} \propto h/H$, where $H$ is the thickness of the disk,
and therefore depends on how well-resolved the vertical disk structure is.
We found very low numerical values, $\alpha_{SS} \sim 10^{-3}$ for the disks
in our models and even lower values in the better resolved central regions 
of the remnant. \\
The whole code is parallelized for shared-memory 
architecture and obtains an excellent speed-up for up to $\sim 100$ processors.
In a typical application with several $10^5$ particles a speed-up of 55 is
obtained on 60 processors.

%
%
We follow the system evolution from an initial separation of $\sim 3 
R_{\rm ns}$, where $R_{\rm ns}$ is the radius of an isolated neutron star,
for approximately 15 ms.
From the chosen initial separation it takes the neutron stars only a few 
milliseconds to merge. They leave behind an extremely massive central 
neutron star ($\sim 2.4$ $\msun$), surrounded by a hot and dense, 
shock-heated inner disk region (with temperatures $T\sim$ 3 MeV, 
densities $\rho \sim 10^{12}$ gcm$^{-3}$ and a mass 
$M_{\rm{disk}}\sim 0.2 \;\msun$) and rapidly expanding debris material.\\
The central neutron star is strongly differentially rotating, most pronounced
in the generic case without initial spin. Since differential rotation
allows the central parts to spin extremely fast without the (slower rotating) 
outer parts of the object reaching the mass shedding limit, a substantially 
higher maximum mass can be stabilized. A recent investigation (Lydford et 
al. 2002) using polytropic equations of state finds values of $(M_r-M_{nr})/
M_{nr}$ up to 1.8 for soft EOSs and for the polytrope closest to 
our nuclear EOS they find $(M_r-M_{nr})/M_{nr}$ up to 0.6 which corresponds
to masses well beyond the total binary mass of 2.8 $\msun$ 
($M_r$ is the maximum mass for a differentially and $M_{nr}$ the maximum 
mass of a non-rotating star). We therefore expect
an extremely massive, hot neutron-star-like object to be formed in the center
of the merger remnant whose lifetime is determined by the time it takes 
to get rid of the rotational support. Although a conclusive
answer to this point cannot be given from the current calculations
(since they are essentially Newtonian and some of the physics ingredients like
the high-density part of the equation of state are to date only poorly known),
we estimate that the neutron star might remain stable 
for many dynamical time scales. This time scale may be long enough to allow
the magnetic seed fields to be amplified to enormous field strengths
($\sim 10^{17}$ G; Thompson and Duncan 1993).
If one assumes magnetic dipole radiation
to drive the system towards black hole formation, even time scales of
months can be easily obtained without stretching the involved parameters beyond
reasonable limits. The exact time scale between the merger event and the
(probable) final black hole formation may depend quite sensitively on
the details of the specific merger event. For a further discussion see
Rosswog \& Davies (2002).\\
The debris around the central object exhibits an 
interesting flow-pattern: material that has previously been centrifugally 
launched into eccentric orbits, and thereby cooled by expansion and 
neutrino-emission, is returning towards the central object. This cool 
($T < 0.5$ MeV), equatorial  inflow  produces a butterfly-shaped 
shock front when it encounters material that is still being shed from the 
central object. In this way a hot flow is driven in vertical direction. 
The resulting disk is very thick with a height comparable to its radial 
extension.

In the present paper we will report on these simulations with a focus
on the neutrino emission that goes along with a neutron star coalescence.

%
%
\begin{table*}
\caption{Summary of the different runs. 
$a_0:$ initial separation; $\nu:$ neutrino physics; 
$T_{sim}:$ simulated duration; $M_1/M_2:$ masses in solar units; 
$\#$ part.:  total particle number}
\begin{flushleft}
\begin{tabular}{cccccccccc} \hline \noalign{\smallskip}
run & spin & ${\rm M_{1}}$ & ${\rm M_{2}}$ & \# part. & $a_{0}$ [km] 
& $\nu$ & $T_{sim}$ [ms] & remark\\ \hline \\
A   & corot. & 1.4 & 1.4 &  207,918 & 48 & no &  10.7 &\\ 
B   & corot. & 1.4 & 1.4 &1,005,582 & 48 & no &  10.8 &\\
C   & irrot. & 1.4 & 1.4 &  383,470 & 48 &yes &  18.3 & gen. case\\
D   & corot. & 1.4 & 1.4 &  207,918 & 48 &yes &  20.2 & lower $\nu$-limit\\
E   & irrot. & 2.0 & 2.0 &  750,000 & 48 &yes &  12.2 & upper $\nu$-limit\\
    &        &     &     &          &    &    &       &\\
F   & corot. & 1.4 & 1.4 &   20,886 &52.5&yes &  15.1 &spur. $\nu$-emission ?\\
\end{tabular}
\end{flushleft}
\label{runs}
\end{table*} 

\section{Neutrino emission from neutron star mergers}\label{neutrino}

Under the conditions of a neutron star merger neutrinos are produced copiously
and they provide the most efficient cooling mechanism for the dense,
shock- and shear-heated neutron star debris. In addition, the related weak 
interactions determine the compositional evolution via the electron fraction 
$Y_e$ that is altered by charged-current reactions such as electron and 
positron captures.
The enormously temperature dependent weak interaction processes can exhibit 
in some parts of the flow very short time scales, 
$|Y_e/\dot{Y}_e| \sim 10^{-6}$ s, which is well below the dynamical time 
scale of a neutron star, $\tau_{\rm dyn}= (G \bar{\rho})^{-1/2} \approx 
2\cdot 10^{-4}$ s,
while they are essentially infinite in other parts of the flow. 
Therefore neither the assumption of an
instantanous beta equilibrium nor frozen $Y_e$ values are justified.
Since in the dense parts of the hot, merged configuration the neutrino 
mean free paths are of the order $\lambda \sim 0.75 \; \rm{m} \;
\left(5\cdot10^{14}\; \rm{gcm}^{-3} / \rho\right)\left(10 \;\rm{MeV}/
T\right)^2$, where $\rho$ is the matter density 
and $T$ is the temperature, the interaction of the 
neutrinos with the ambient matter has to be accounted for.
Here and in the rest of the paper we measure temperatures in energy units,
i.e. $k_B=1$.
A full Boltzmann neutrino transport in the context of the three-dimensional 
modelling of the event is beyond the current state-of-the-art and 
computational resources. 
But since the simulated physical time scales are of the order of 10 
milliseconds and neutrino momentum transfer is expected to be unimportant 
we consider a detailed neutrino leakage to be an important step towards 
reliable physical models of the event.\\
We consider three neutrino flavours: electron neutrinos, $\nu_e$, electron 
anti-neutrinos, $\bar{\nu}_e$, and the heavy-lepton neutrinos, $\nu_{\mu},
\bar{\nu}_{\mu},\nu_{\tau},\bar{\nu}_{\tau}$, which are collectively 
referred to as $\nu_x$. The basic idea of our leakage scheme is to 
provide a physical limit via diffusion rates. This guarantees the limitation
of the neutrino production to the amount that is able to stream away.
In the opaque regime the neutrinos therefore escape only on a diffusion
time scale and in the transparent regions they leave their production site 
essentially without any further interaction with the surrounding matter.\\ 
The dominant neutrino processes in our context are the charged-current 
lepton capture reactions on nucleons,
electron capture (EC)
\begin{equation} 
     e^- + p \rightarrow n + \nu_e \label{EC} 
\end{equation}
and positron capture (PC)
\begin{equation}
     e^+ + n \rightarrow p + \bar{\nu}_e \label{PC}
\end{equation}
which produce electron flavour neutrinos
and the ``thermal'', pair producing reactions,
pair annihilation
\begin{equation} 
     e^- + e^+ \rightarrow \nu_i + \bar{\nu}_i \label{pair} 
\end{equation}
and plasmon decay
\begin{equation} 
     \gamma \rightarrow \nu_i + \bar{\nu}_i \label{plasma}, 
\end{equation}
which produce neutrinos and anti-neutrinos of all flavours, 
$\nu_i$ and $\bar{\nu}_i$. 
The latter process dominates in the strongly electron-degenerate regime. 
We disregard electron captures onto nuclei since these reactions would 
require detailed information about the 
nuclear shell structure which is not available for these nuclei.
But these captures are not expected to be important in our case, since 
the regimes where the dominant neutrino emission takes place are almost 
completely photodisintegrated (see below). We further neglect neutral-current
nucleon-nucleon bremsstrahlung as neutrino production process. This process
has recently received attention in the supernova context (Thompson, Burrows
and Horvath 2000). It may be possible that this process is locally important,
but recent investigations (Keil et al. 2002) including this and other 
reactions for the supernova case only
found overall changes of the order 10 \% . Since accurate emission
rates are difficult to obtain (due to the poor knowledge of the 
nucleon-nucleon potential and due to uncertainties in the magnitude of 
many-body effects) and we do not expect effects larger than the uncertainties
 inherent in our leakage scheme, we decided to ignore this process.\\
To determine the number and energy diffusion rates based on neutrino opacities
we take  into account the scattering off nucleons,
   \begin{equation} \nu_i + \{n, p\} \rightarrow \nu_i + \{n, p\},
   \label{nuc_scat}
   \end{equation}
coherent neutrino nucleus scattering,
   	\begin{equation}
        \nu_i + A \rightarrow \nu_i + A \label{A_scat},
   	\end{equation} and
neutrino absorption by free nucleons,
	\begin{eqnarray}
	\nu_e + n \rightarrow p + e^-\\
	\bar{\nu}_e + p \rightarrow n + e^+ .
	\end{eqnarray}
For the details of the implementation of the reactions and the leakage scheme
we refer to the Appendix.

\subsection{Total luminosities, mean energies}
\begin{figure}
\hspace*{-0.5cm}\psfig{file=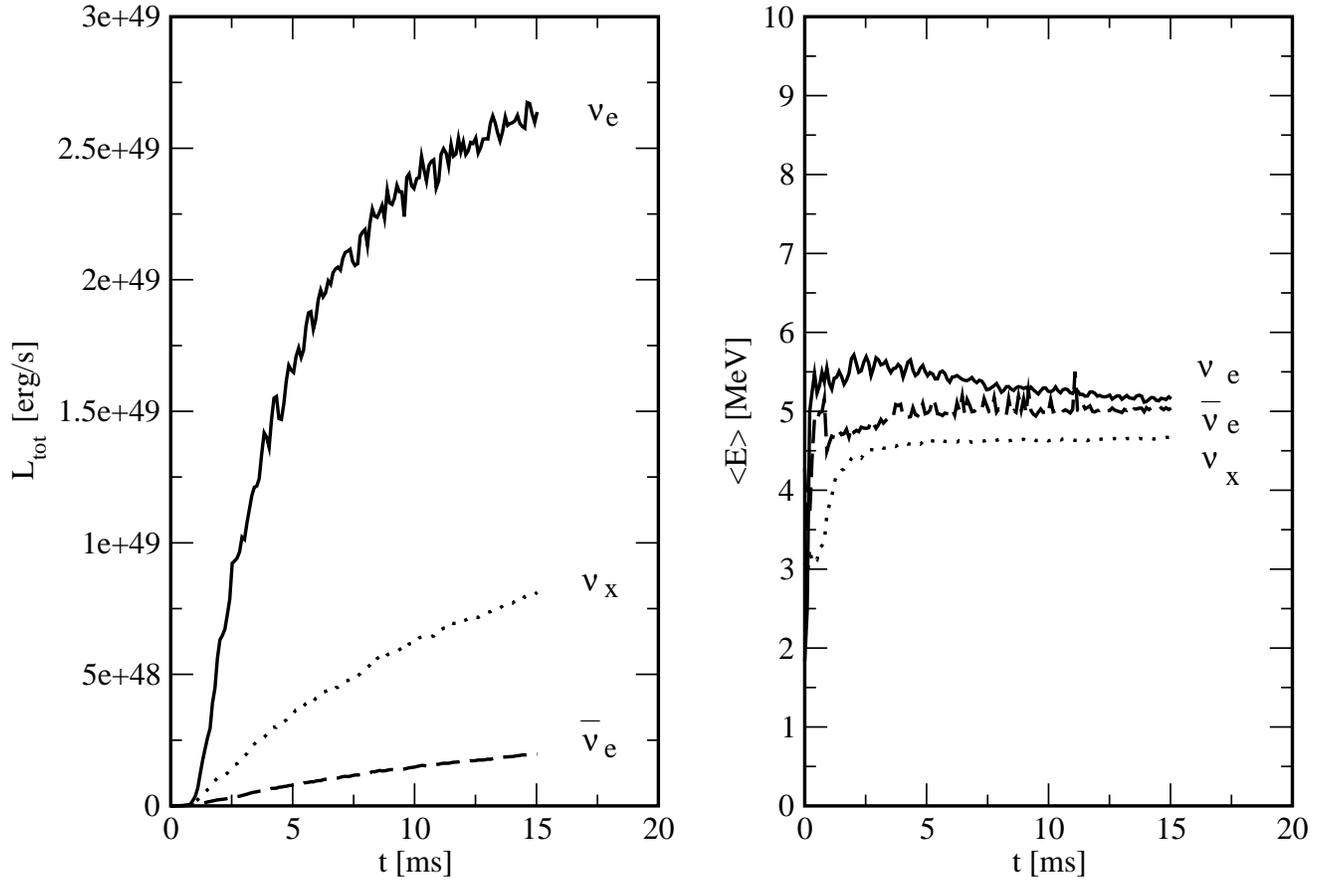,width=15cm,angle=-90}
    \caption{\label{spur_em}Testing for spurious neutrino emission: 
             Shown are the luminosities of the various neutrino species (left)
             and the corresponding mean energies.
             After 15 ms, corresponding to $\sim$ 50 neutron star dynamical 
             time scales, the total neutrino luminosity levels off four orders 
             of magnitude below the emission of the full merger.}
\end{figure}

Neutron stars are expected to heat up during inspiral by tidal interaction 
to temperatures of the order $10^8$ K (Lai 1994).
At these temperatures no significant neutrino emission will occur.
In order to test for the amount of ``spurious'' emission of
neutrinos in our simulation due to the unavoidable numerical heat up 
of the completely degenerate stars, we perform a test run (a listing of
the different runs is provided in Table \ref{runs}). 
We prepare a corotating equilibrium binary configuration
just outside the last stable orbit by relaxing the two neutron stars in
their mutual gravitational field. Subsequently we follow their dynamical
evolution for approximately 50 neutron star dynamical time scales
while they revolve on perfectly circular orbits around their 
common center of mass. Only self-gravity and hydrodynamic forces are 
considered,  no initial radial velocities are applied
and the gravitational wave backreaction forces are switched off.
We find that the total neutrino luminosity reaches a stationary level of  
$\sim 3 \cdot 10^{49}$ erg s$^{-1}$, see Fig. \ref{spur_em}, which is 
four orders of magnitude below the peak luminosities of the full merger 
calculation and therefore completely negligible.

\begin{figure}
   \hspace*{-0.5cm}\psfig{file=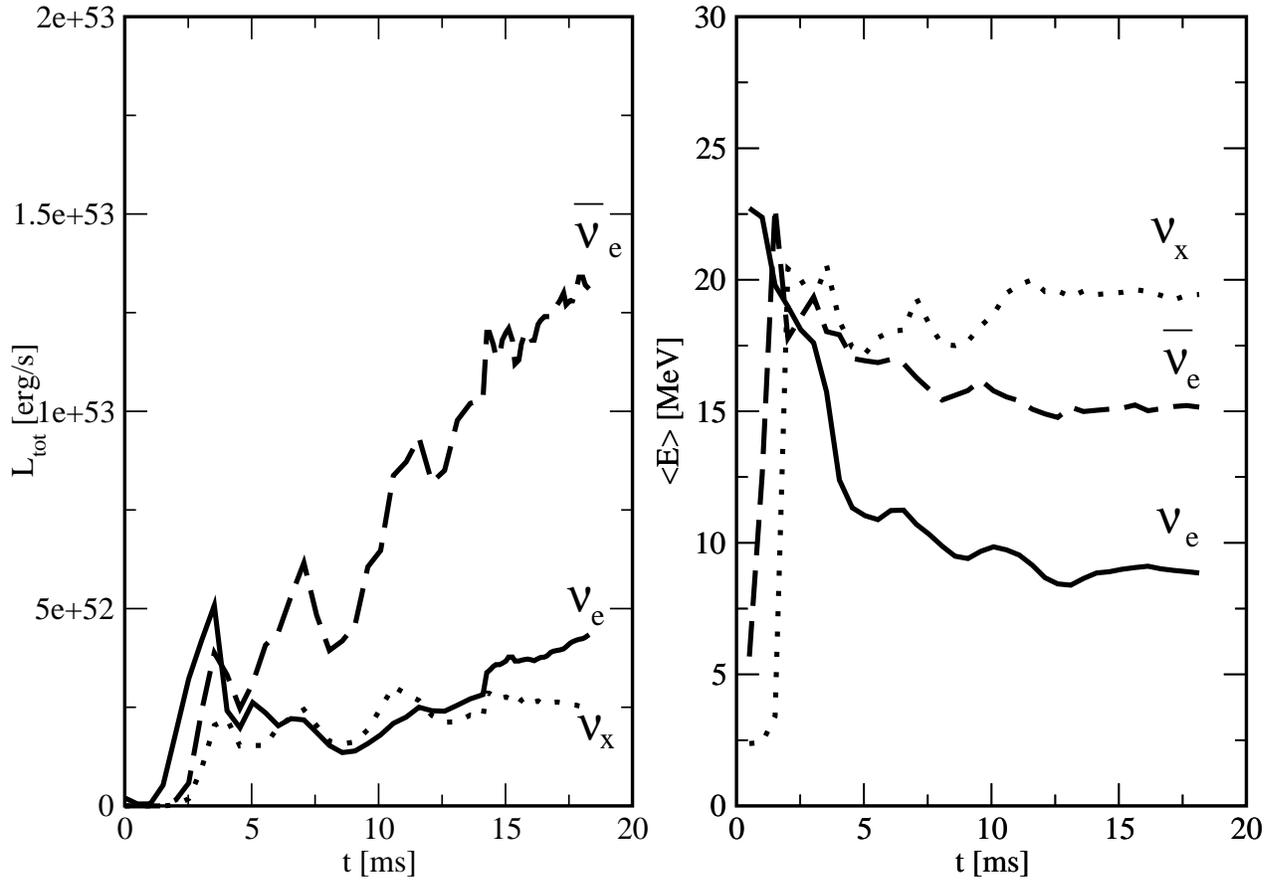,width=15cm,angle=-90}	 
   \caption{\label{nu_lum_ir200}Run C (no spins, 2 x 1.4 $\msun$): 
            the left panel shows the luminosities 
            (in ergs/s) of the different neutrino flavours. The right panel
            gives the corresponding mean energies. We regard this to be the 
            generic case.}
\end{figure}

\begin{figure}
\hspace*{-0.5cm}\psfig{file=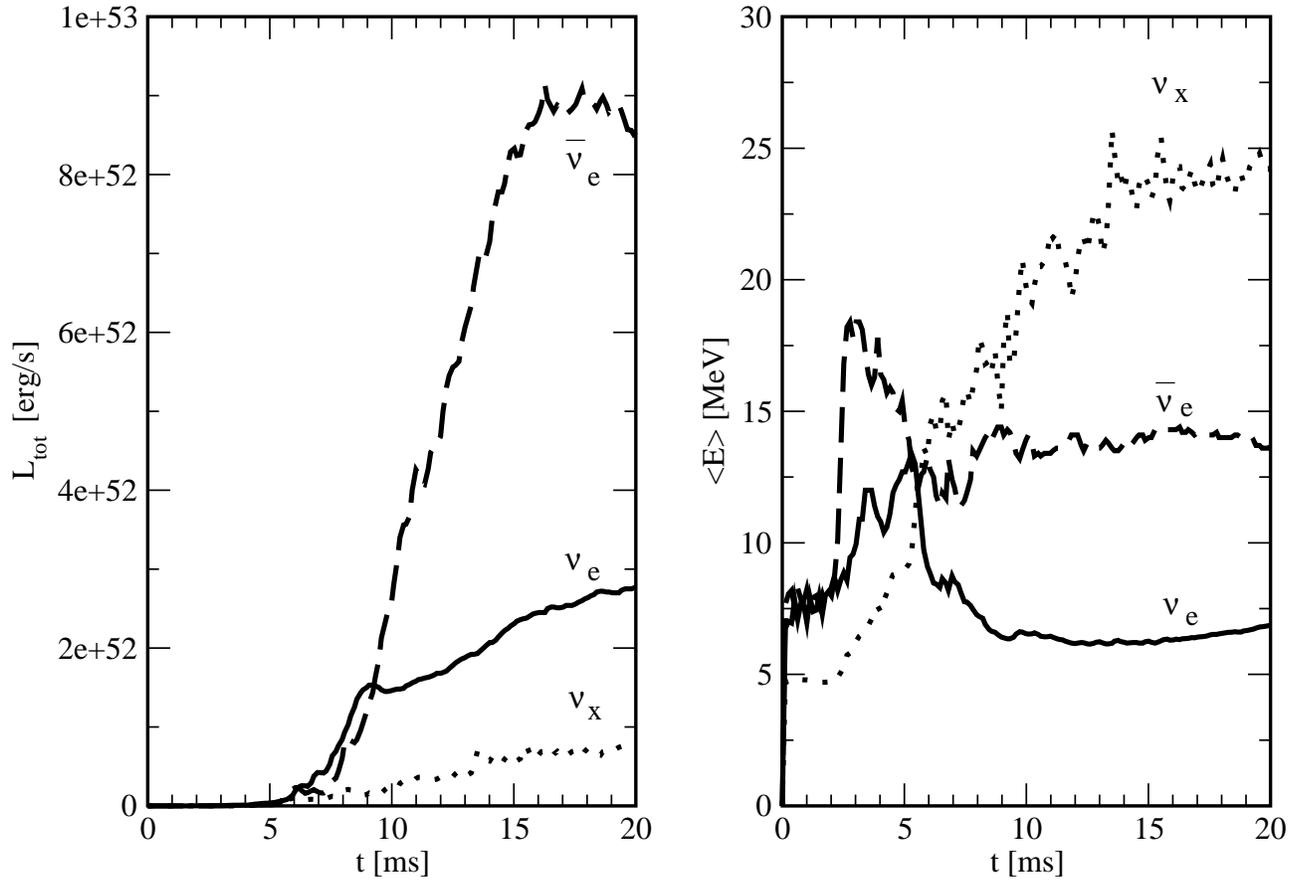,width=15cm,angle=-90}
			
    \caption{\label{nu_lum_corot}Run D (corotation, 2 x 1.4 $\msun$): 
            the left panel shows the luminosities 
            (in ergs/s) of the different neutrino flavours. The right panel
            gives the corresponding mean energies.}
\end{figure}

\begin{figure}
\hspace*{-0.5cm}\psfig{file=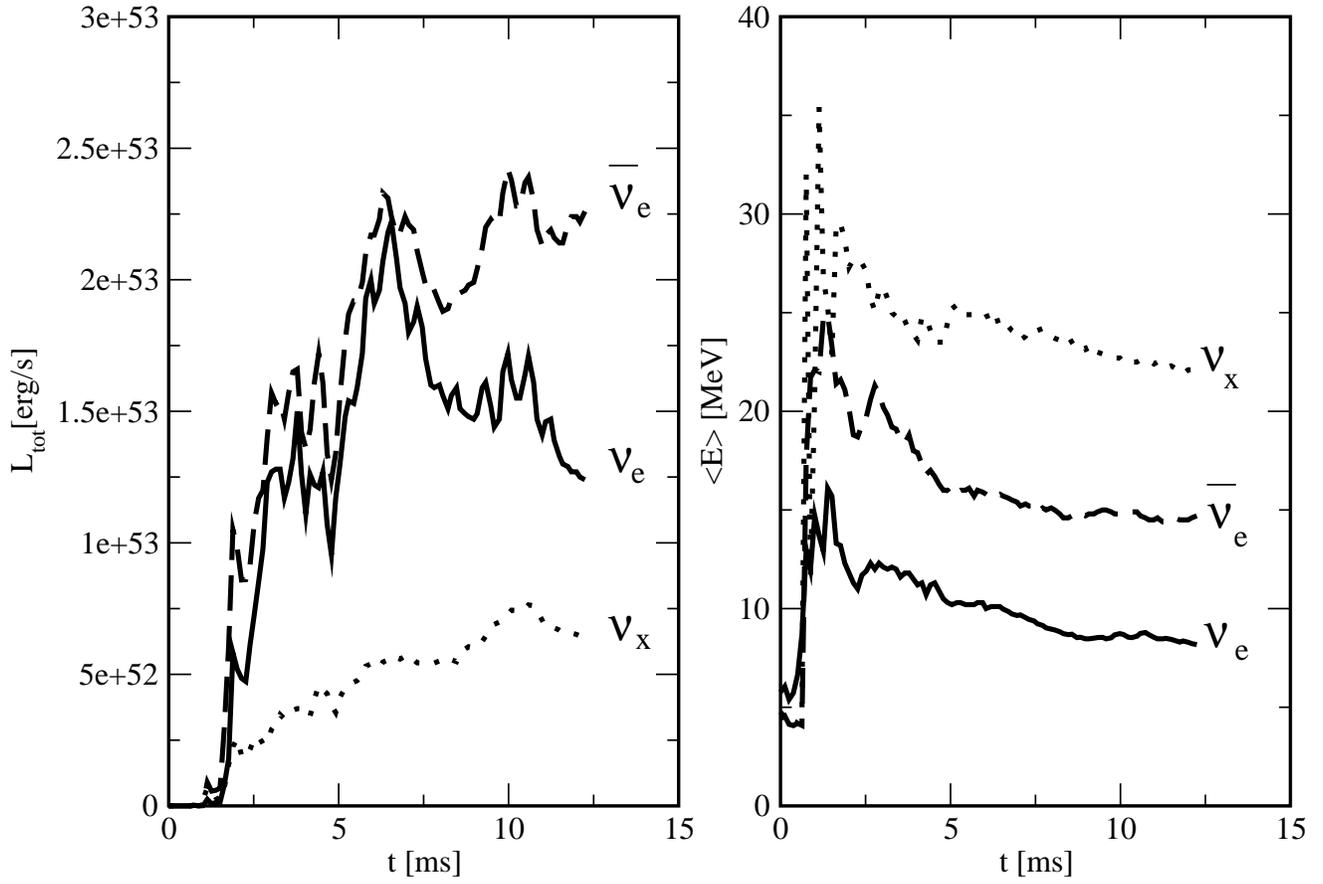,width=15cm,angle=-90}
    \caption{\label{nu_lum_ir700}Run E (no spins, 2 x 2.0 $\msun$): 
            the left panel shows the luminosities 
            (in ergs/s) of the different neutrino flavours. The right panel
            gives the corresponding mean energies. We regard this to be an
            upper limit for the neutrino emission.}
\end{figure}

\begin{figure}
    \centerline{\psfig{file=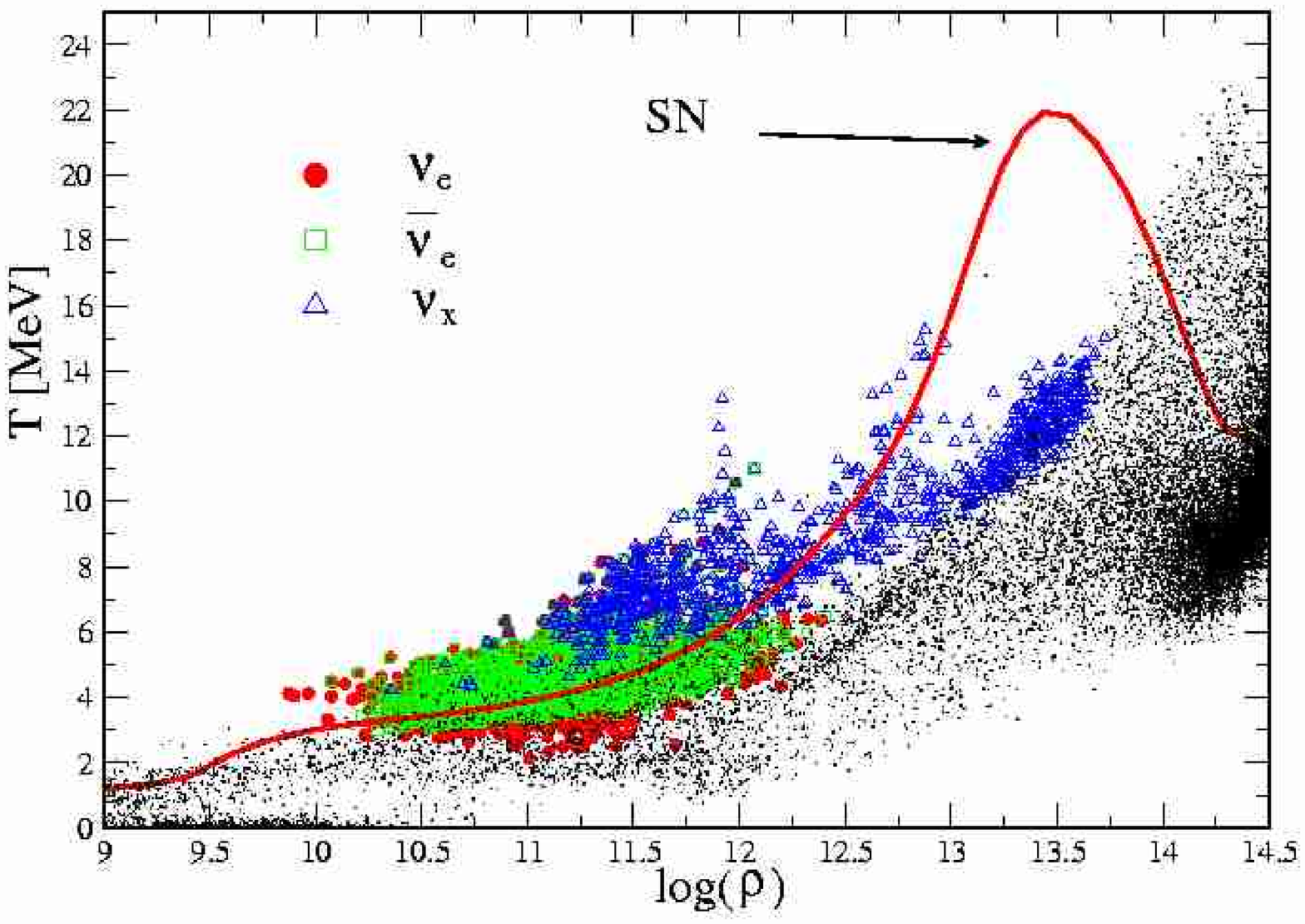,width=10cm,angle=-90}}
	\centerline{\psfig{file=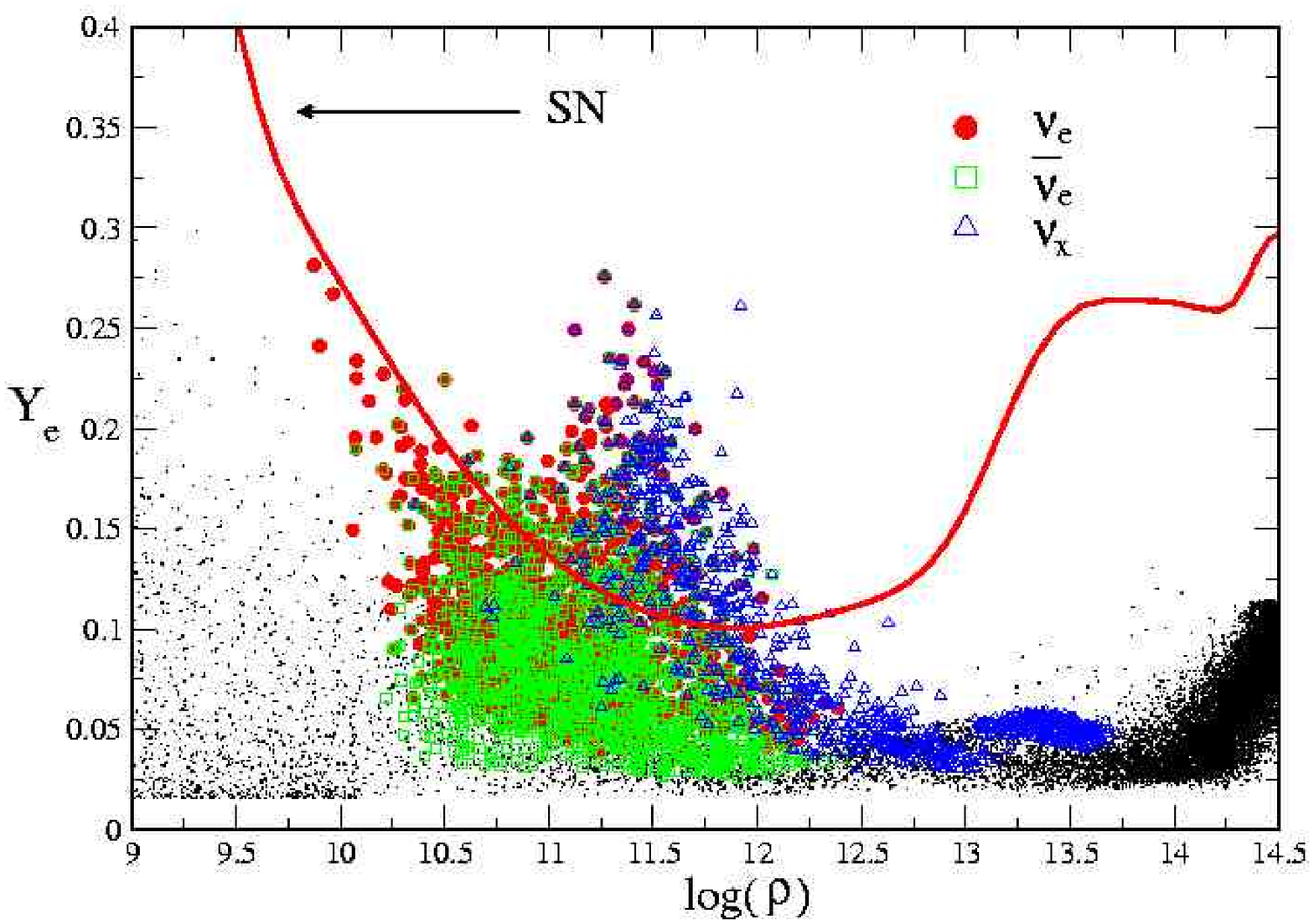,width=10cm,angle=-90}}
    \caption{\label{lrho_T} Shown are the SPH-particle distributions in the
log($\rho)-T$- (upper panel) and the log($\rho)-Y_e$-plane (lower panel) of 
the generic case, run C (2 x 1.4 $\msun$, no initial spin; every 20th particle
is displayed as a dot). For comparison with the supernova (SN) case we 
show the conditions of a collapsed 13 $\msun$ star, 100 ms after bounce 
as a thick line. The particles with the highest luminosities are also shown:
filled circles for the $\nu_e$, squares for the $\bar{\nu}_e$ and the 
triangles refer to the $\nu_x$.}

\end{figure}

The overall neutrino emission properties of the full merger calculations
are shown in Figs. \ref{nu_lum_ir200}
to \ref{nu_lum_ir700}. The total neutrino luminosities (left panels) are 
calculated summing up all particle contributions and the rms energies
for each neutrino flavour are calculated according to eq. (\ref{mean_E}). 
For the corotating case, run D, substantial neutrino emission sets in 
later than in the cases without initial spin.
The explanation for this is twofold: on the one hand this run starts out
from numerically exact initial conditions while the non-rotating cases, run
C and E, suffer an accelerated inspiral due to the start with initially
spherical stars. On the other hand, neutrino emission only becomes 
important once the thick torus around the central high-density part has 
formed.
Again this process takes longer for the corotating case since 
(due to the larger initial angular momentum) the torus-forming matter is
initially launched into wider orbits. The average neutrino energies
reach peak values soon after the stars have first come into contact.
The reason for this is that the neutrinos from the hot debris material 
can, at this stage, escape without having to pass through any optically 
thick matter.\\
The total neutrino emission seems to have reached a roughly stationary 
level (except for maybe run C) by the end of the simulation. The total 
luminosities range from $\sim 10^{53}$ erg/s for the smoothly merging 
corotating case over $\sim 2 \cdot 10^{53}$ erg/s for the irrotational
case with twice 1.4 $\msun$ to $\sim 4 \cdot 10^{53}$ erg/s for our extreme 
case with 2 x 2.0 $\msun$ and no initial spins. We regard run D as a lower 
limit which is unlikely to occur in nature (Bildsten \& Cutler 1992, 
Kochanek 1992)
and run C as the generic case since the observed
neutron star binary systems have masses close to 1.4 $\msun$ (Thorsett \& 
Chakrabarti 1999)
and are expected to have a very slow individual spin at the merger stage. 
The extreme case run E has been performed in order to explore the upper 
limit on the neutrino emission from the merger event.\\
The mean energies are $\sim$ 8 MeV for the electron-type neutrinos.
This is below the typical values found in core collapse supernovae.
The material in the nascent protoneutron star has first to deleptonize
on a neutrino diffusion time scale before the electron fractions are
as low as in the neutron star merger event, where beta-equilibrium
has been established in the individual neutron stars long before the 
coalescence. Hence, the material in the supernova is more electron degenerate
at comparable densities.  Electrons are therefore captured from higher Fermi 
energies and produce electron neutrinos with a harder spectrum in the 
supernova case.
The situation is different for the electron antineutrinos. We find rms energies
around \( \sim 15 \) MeV, quite comparable to rms energies in the
supernova case. The lower electron degeneracy in the neutron star
merger favours the population of positrons, whose chemical potential
has to balance the electron chemical potential because of pair equilibrium.
The high positron abundance in combination with the neutron rich matter
leads to more positron capture events on free neutrons than in the
supernova. This results in higher electron anti-neutrino luminosities.
The electron anti-neutrino luminosity from the remnant reaches
up to $\sim 1.5 \cdot 10^{53}$ erg/s, it provides the main cooling 
mechanism of the hot accretion disk. The heavy lepton neutrinos reach rms 
energies of \( \sim 20 \) to \( \sim 25 \) MeV. This is comparable 
to the supernova rms energies.  Their luminosity, however, tends to be 
smaller in the neutron star merger case because the temperatures in 
the high density regimes, where the heavy lepton neutrinos emerge,
are manifestly lower (see below).\\
To characterize the physical conditions of the emission region of each 
neutrino flavour we calculate average quantities, $\tilde{X}_{\nu_i}$, 
weighted by the $\nu_i$-number production rates per particle, 
$\tilde{R}^{ef}_{\nu_i,j}$, where $X$ stands for $\rho,T,Y_e$ and $\mu_e$, 
given by
\begin{equation}
\tilde{X}_{\nu_i}= 
\frac{\sum_j \tilde{R}^{ef}_{\nu_i,j} \; X_j}{\sum_j \tilde{R}^{ef}_{\nu_i,j}}\label{X}.
\end{equation}
These quantities are displayed in Table 
\ref{typical_properties}. Electron neutrinos and anti-neutrinos are emitted 
under similar conditions, typically at densities around $10^{12.5}$ gcm$^{-3}$ 
, temperatures of 4-5 MeV (anti-neutrinos at slightly higher 
values) and a $Y_e$ below 0.1. The heavy lepton neutrinos are emitted at 
substantially higher densities (log$(\rho)\approx 13$) and 
temperatures ($T \approx$ 9 MeV).\\ 
Figure \ref{lrho_T} shows a comparison between the conditions encountered 
in neutron star mergers and those of SNe.
In the upper panel we show SPH-particle densities and temperatures 
(black dots, every 20th particle is shown) for our generic run C. 
Particles with peak luminosities are indicated with special symbols
(these {\em peak} values are not to be confused with the {\em average} 
properties mentioned above).
Filled circles indicate particles which
emit $\nu_e$ at a luminosity in excess of 10 $\%$ of the maximum particle 
$\nu_e$-luminosity, squares mark the corresponding particles for $\bar{\nu}_e$
emission and triangles refer to $\nu_x$. Due to the steeper temperature
dependence ($Q_{\nu_x} \propto T^9$) a lower threshold (3.5 $\%$) has
been chosen for the $\nu_x$ in order to display roughly the same number of 
particles. The corresponding
plot for $Y_e$ as a function of $\log(\rho)$ is shown in the second panel of 
Figure \ref{lrho_T}.  We
compare the state of these  fluid elements to the state of
fluid elements in a simulated postbounce evolution of the core of
a \( 13 \) M\( _{\odot } \) progenitor star. Due to the limitation
to spherical symmetry in this simulation with Boltzmann neutrino transport
(Liebend\"orfer et al., 2002), the fluid elements form
 a solid line. In the supernova case, at \( 100 \) ms after
bounce, we find the peak emission at densities of \( 10^{10.6} \),
\( 10^{11.3} \), and \( 10^{12.6} \) g/cm\( ^{3} \) respectively.
This is in  agreement with the high emission regions identified
in the neutron star merger for the heavy lepton neutrinos and the
electron anti-neutrinos. The peak emission of electron neutrinos in the
neutron star merger appears to occur at slightly higher densities
(\( \sim 10^{11} \) g/cm\( ^{3} \)). We attribute this to the less
pronounced compression and deleptonization of infalling matter at
low densities in the rotating accretion disk if compared to the failed
explosion of a non-rotational supernova simulation. 
The heavy lepton neutrinos stem in both SN and neutron star merger
from similar densities.\\
To analyze the importance of the e$^+$/e$^-$-capture reactions, eqs.
(\ref{EC}) and (\ref{PC}), versus
the pair producing reactions eqs. (\ref{pair}) and (\ref{plasma})
we perform a post-processing experiment.
We take one time-slice of our generic run, run C, at t= 14.1 ms and use 
the pair and plasma neutrino reactions as the only emission processes
(i.e. the capture reactions are artificially switched off). In this case
the luminosity in electron-type (anti-) neutrinos is only $\sim $ 10\% of the
previous values, indicating that a major contribution stems from the
lepton capture reactions.

\begin{table*}
\caption{Typical properties of emission region, densities are in gcm$^{-3}$,
temperatures and chemical potentials in MeV.}
\begin{flushleft}
\begin{tabular}{cccccccccccccc} \hline \noalign{\smallskip}
run&$log(\tilde{\rho})_{\nu_e}$& $log(\tilde{\rho})_{\bar{\nu}_e}$&$log(\tilde{\rho})_{\nu_x}$&
$\tilde{T}_{\nu_e}$&$\tilde{T}_{\bar{\nu}_e}$&$\tilde{T}_{\nu_x}$&
$\tilde{Y}_{e,\nu_e}$&$\tilde{Y}_{e,\bar{\nu}_e}$&$\tilde{Y}_{e,\nu_x}$&
$\tilde{\mu}_{e,\nu_e}$&$\tilde{\mu}_{e,\bar{\nu}_e}$&$\tilde{\mu}_{e,\nu_x}$&\\ \hline \\
C   &12.6 & 12.6 & 13.2 & 4.2 & 5.5 & 8.9 & 0.072 & 0.072& 0.13& 12.1 & 13.2 & 23.7 & \\
D   &12.4 & 12.2 & 13.5 & 4.1 & 5.0 & 6.6 & 0.083 & 0.095& 0.10& 11.6 & 13.0 & 35.1 & \\
E   &12.1 & 12.4 & 12.9 & 5.0 & 5.7 & 9.1 & 0.140 & 0.085& 0.12&  9.6 & 13.0 & 22.1 & \\
\end{tabular}
\end{flushleft}
\label{typical_properties}
\end{table*}

\subsection{Emission geometry: disk versus central object}

\begin{figure*}
\centerline{\psfig{file=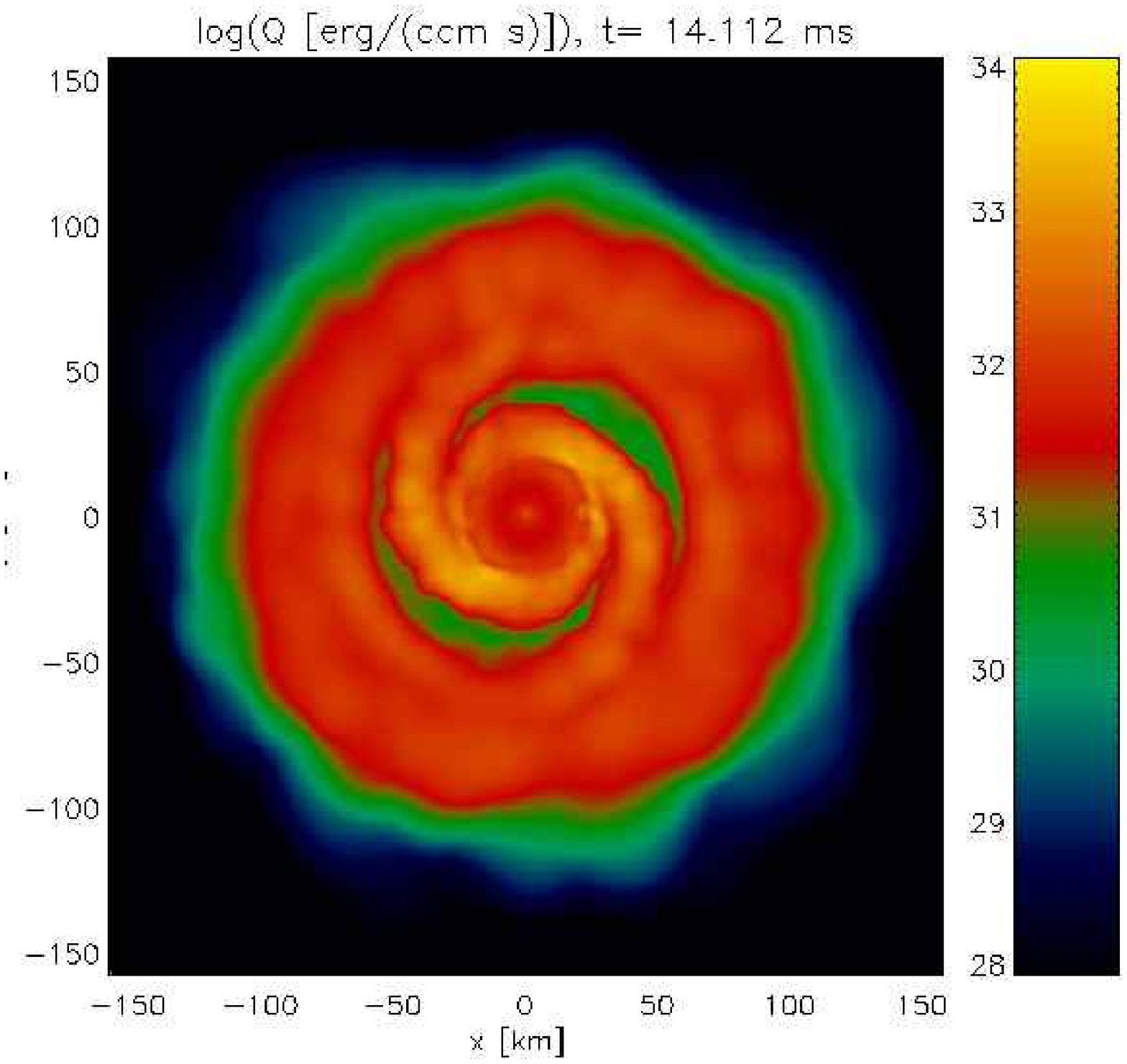,width=7cm,angle=0}
\hspace{1cm}\psfig{file=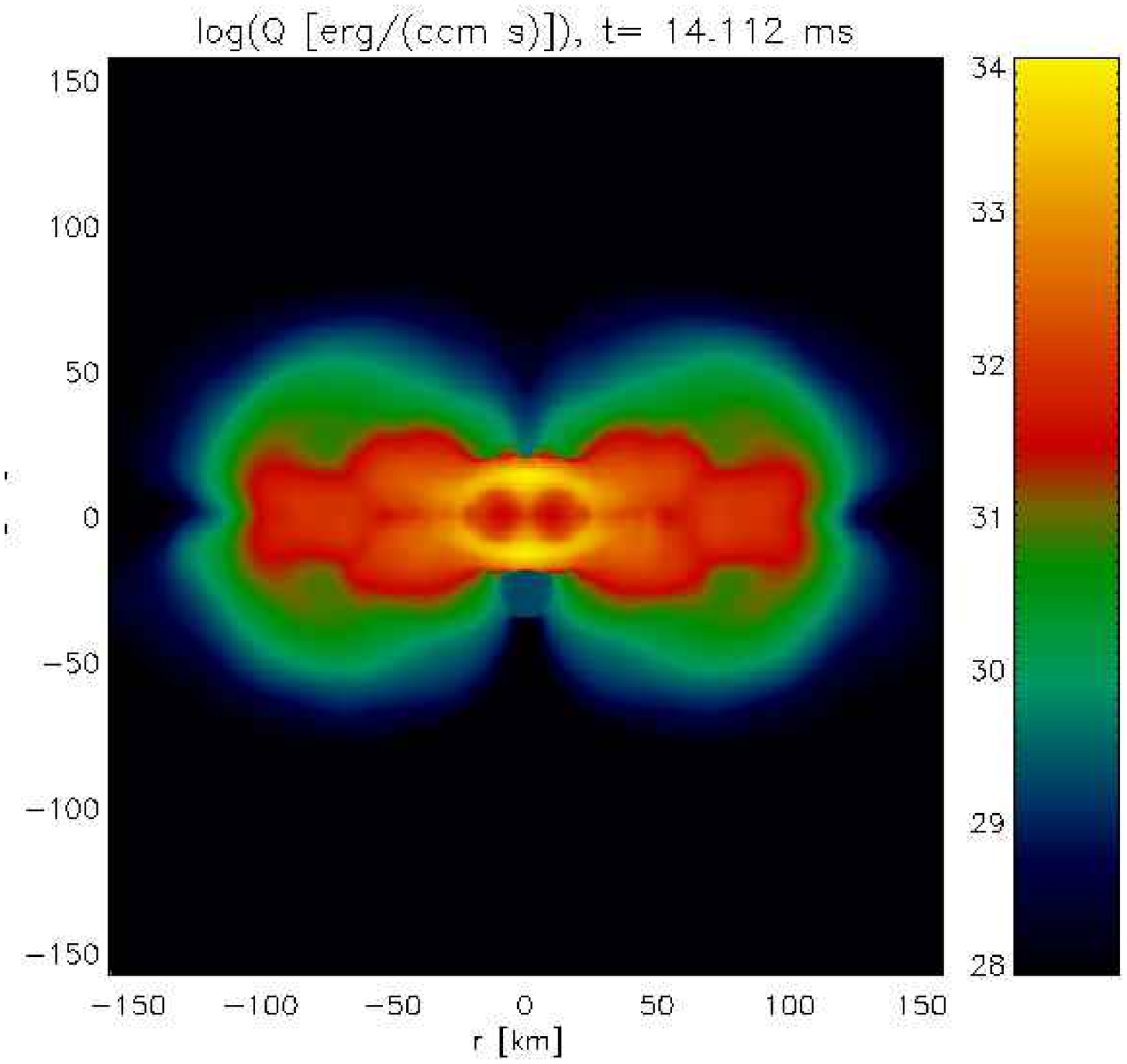,width=7cm,angle=0}}
\centerline{\psfig{file=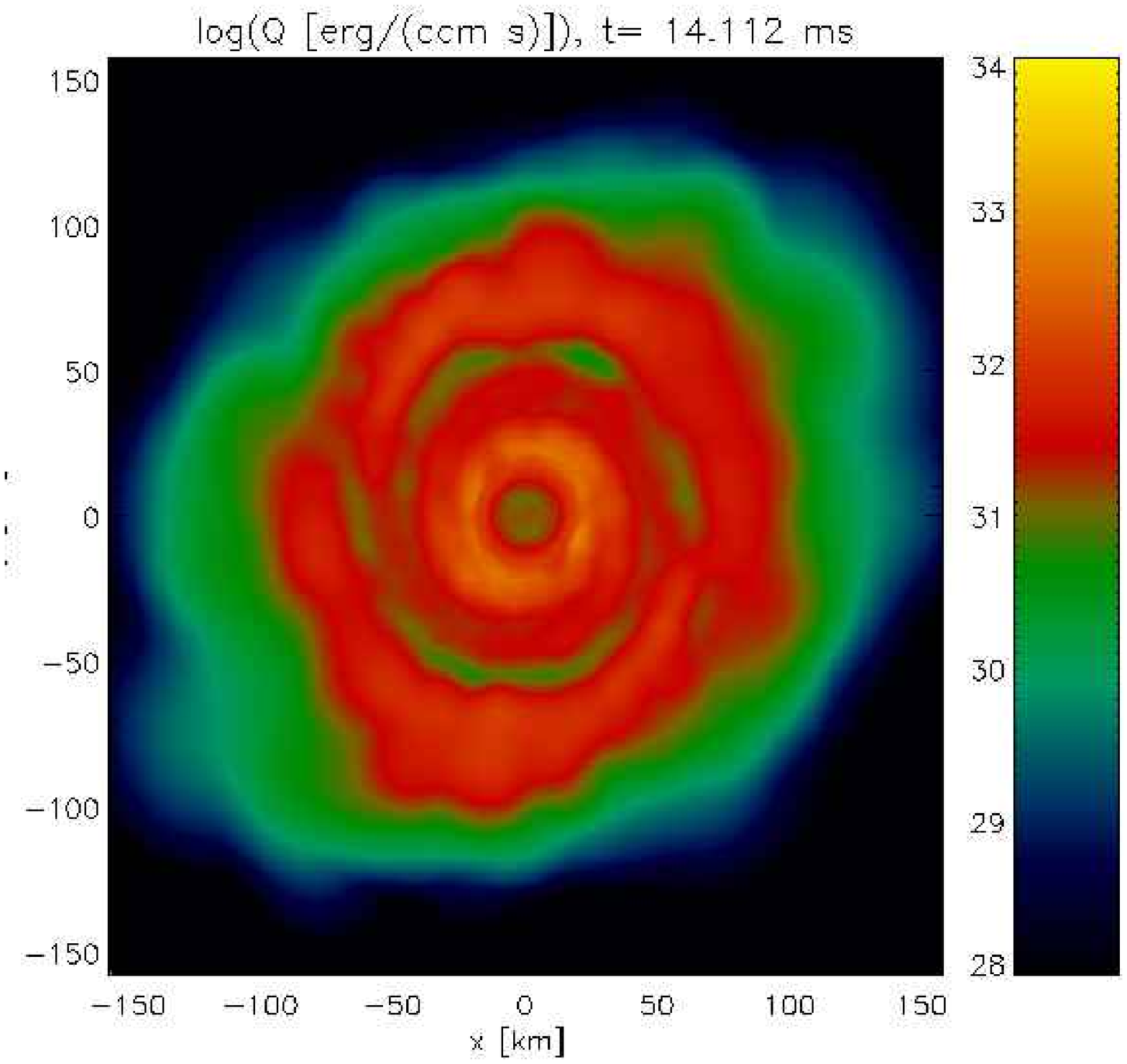,width=7cm,angle=0}
\hspace{1cm}\psfig{file=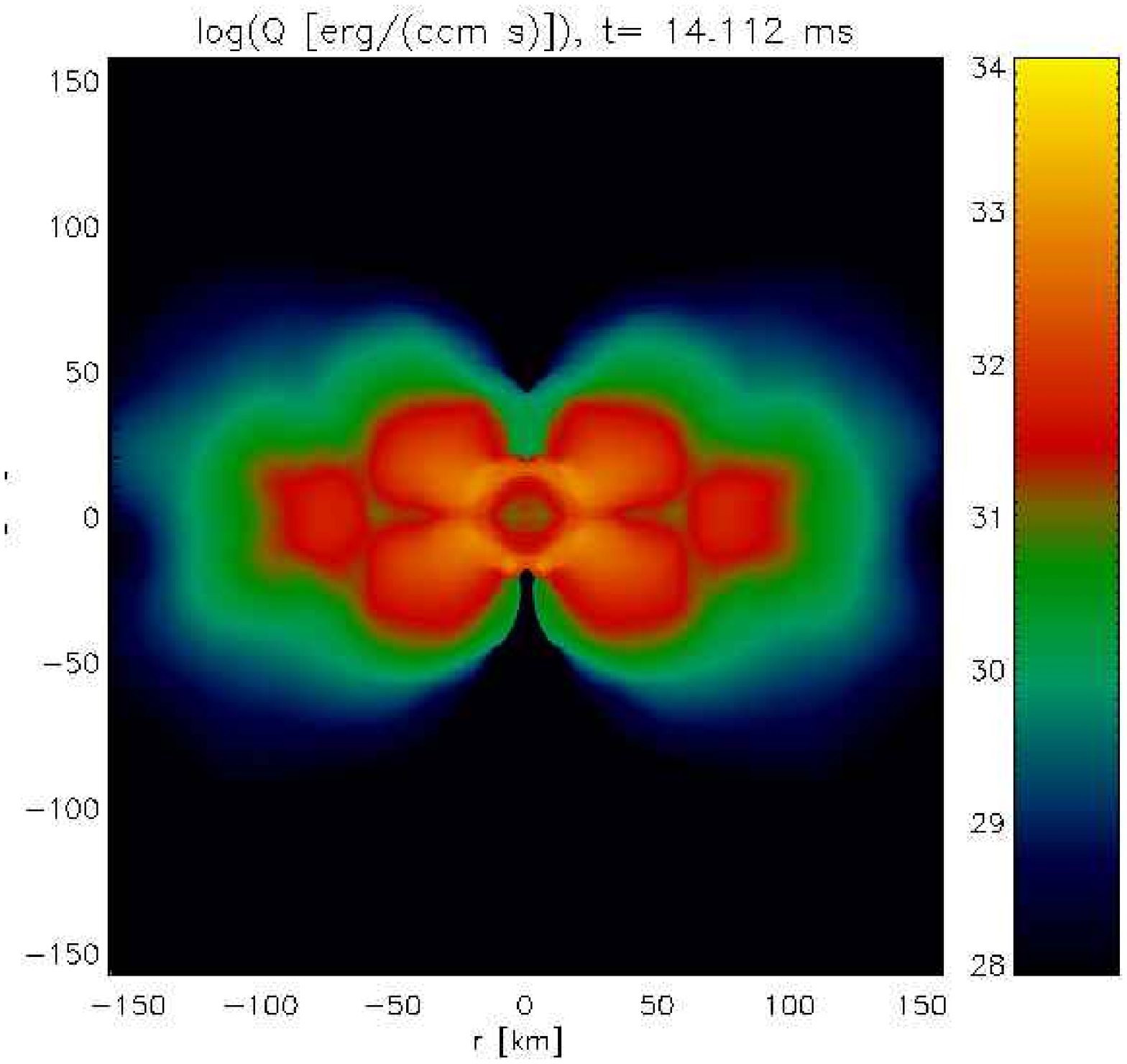,width=7cm,angle=0}}
\centerline{\psfig{file=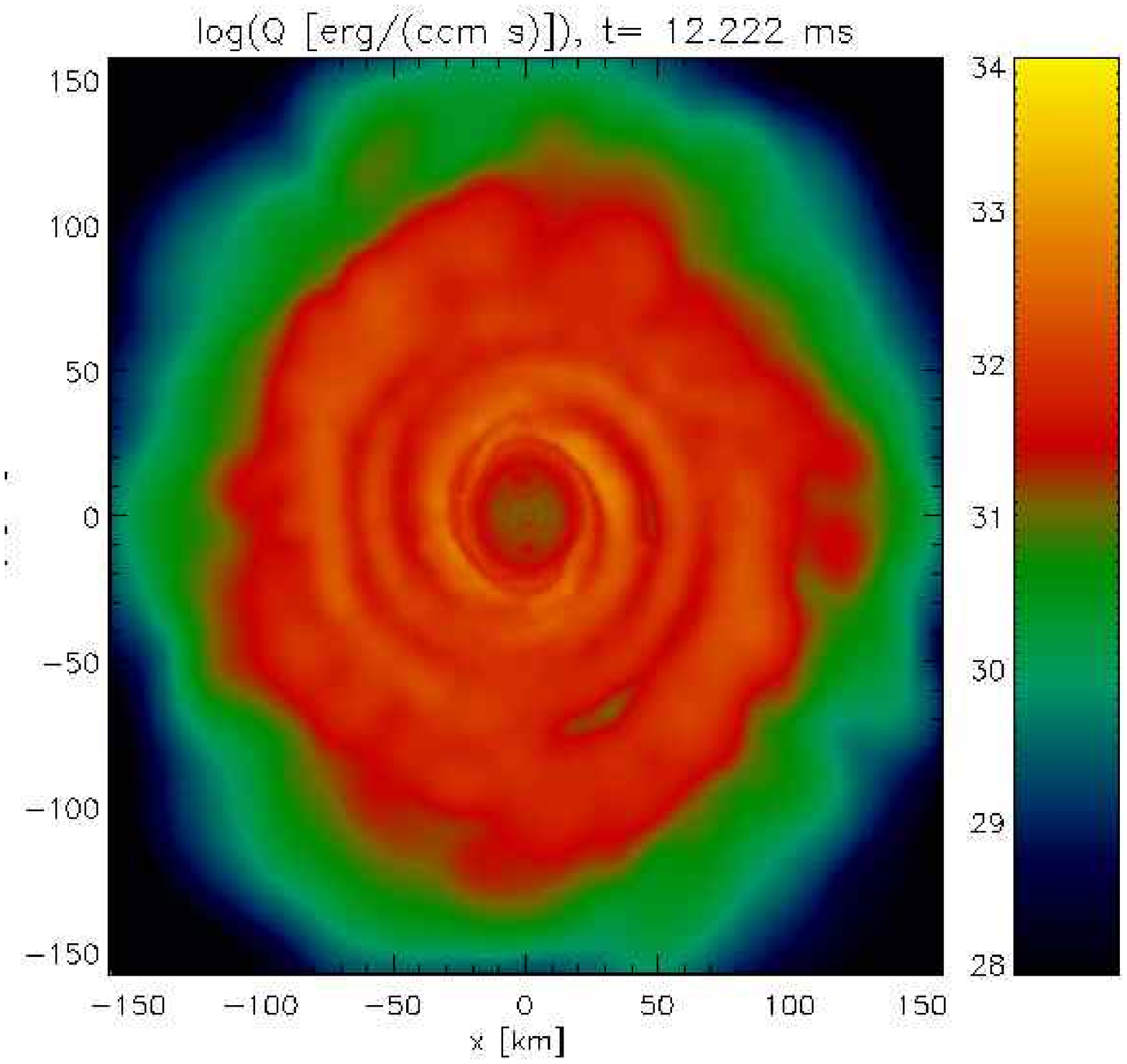,width=7cm,angle=0}
\hspace{1cm}\psfig{file=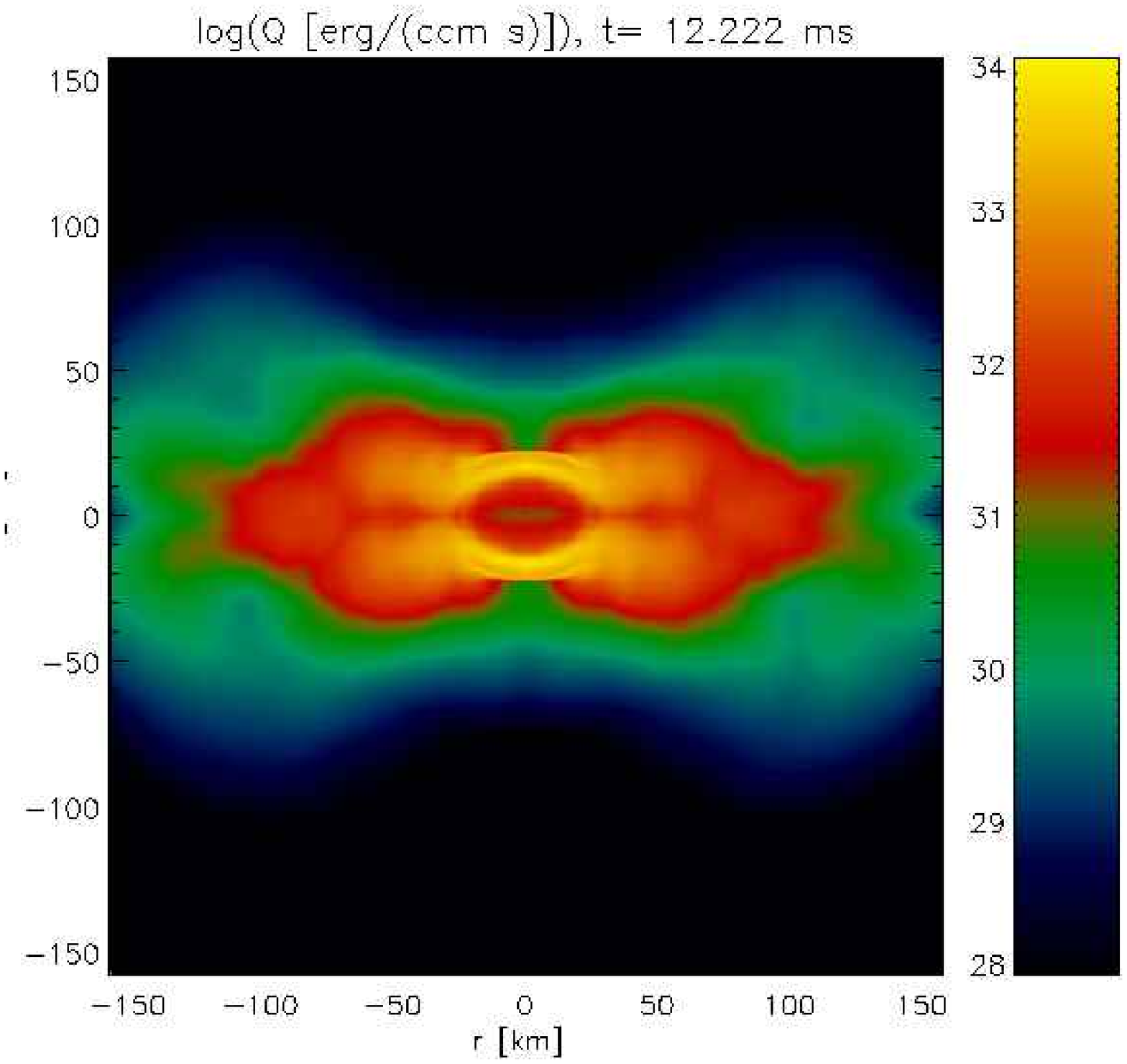,width=7cm,angle=0}}

    \caption{\label{em_geom}Geometry of the neutrino emission: 
             the left column shows the total neutrino energy per
             time and volume in the orbital plane, the right column 
             displays the vertical emission geometry. 
	     The upper two panels correspond to run C (2 x 1.4 $\msun$, 
             no spins), the middle panels to run D (2 x 1.4 $\msun$,
             corotation) and the lowest panels to run E (2 x 2.0
             $\msun$, no spins).
	     The contribution of the central object is negligible.}
\end{figure*}

In Fig. \ref{em_geom} we plot the neutrino energy (sum of all flavours)
per time and volume for run C (upper two panels), run D (intermediate two
panels) and run E (the two lower panels). The left column of panels shows
the emission in the orbital plane, while the vertical emission geometry
(azimuthally averaged) is displayed in the right column. Note that the 
emission per time and volume from the hot, but extremely dense central 
objects is completely
negligible, roughly two orders of magnitude lower than that coming from
the most luminous parts of the disk. In paper I we had mentioned the 
butterfly-shaped temperature distribution in the XZ-plane that results
from cool inflow being shock-heated hitting the inner parts of the disk,
see Figure 15 in paper I.
This pattern is also reflected in the neutrino emission geometry, see
right column in Fig. \ref{em_geom}.\\

\subsection{Opacity sources: Importance of heavy nuclei}

\begin{figure}
\vspace*{0cm}
\centerline{\psfig{file=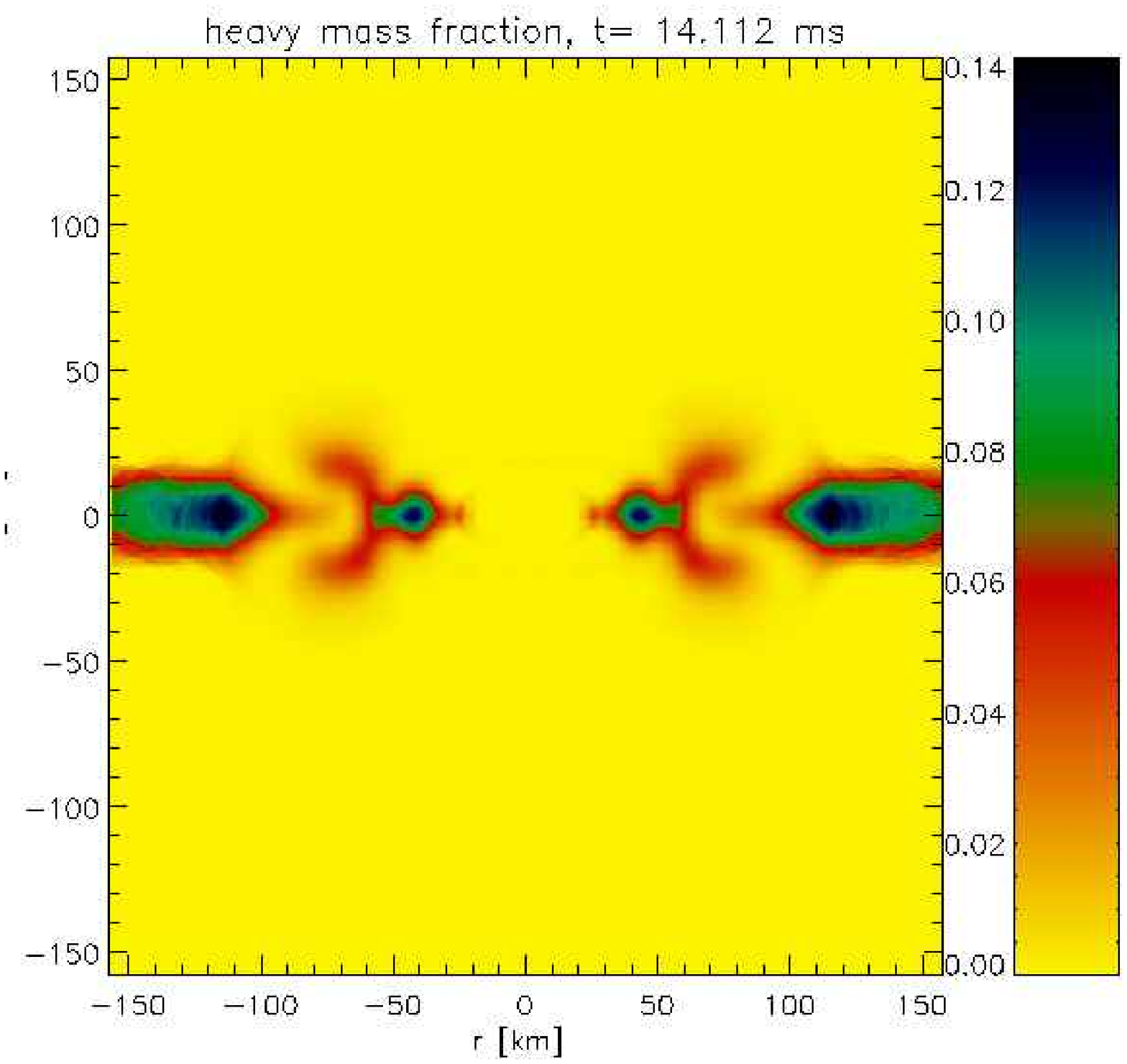,width=7cm,angle=0}}
\centerline{\psfig{file=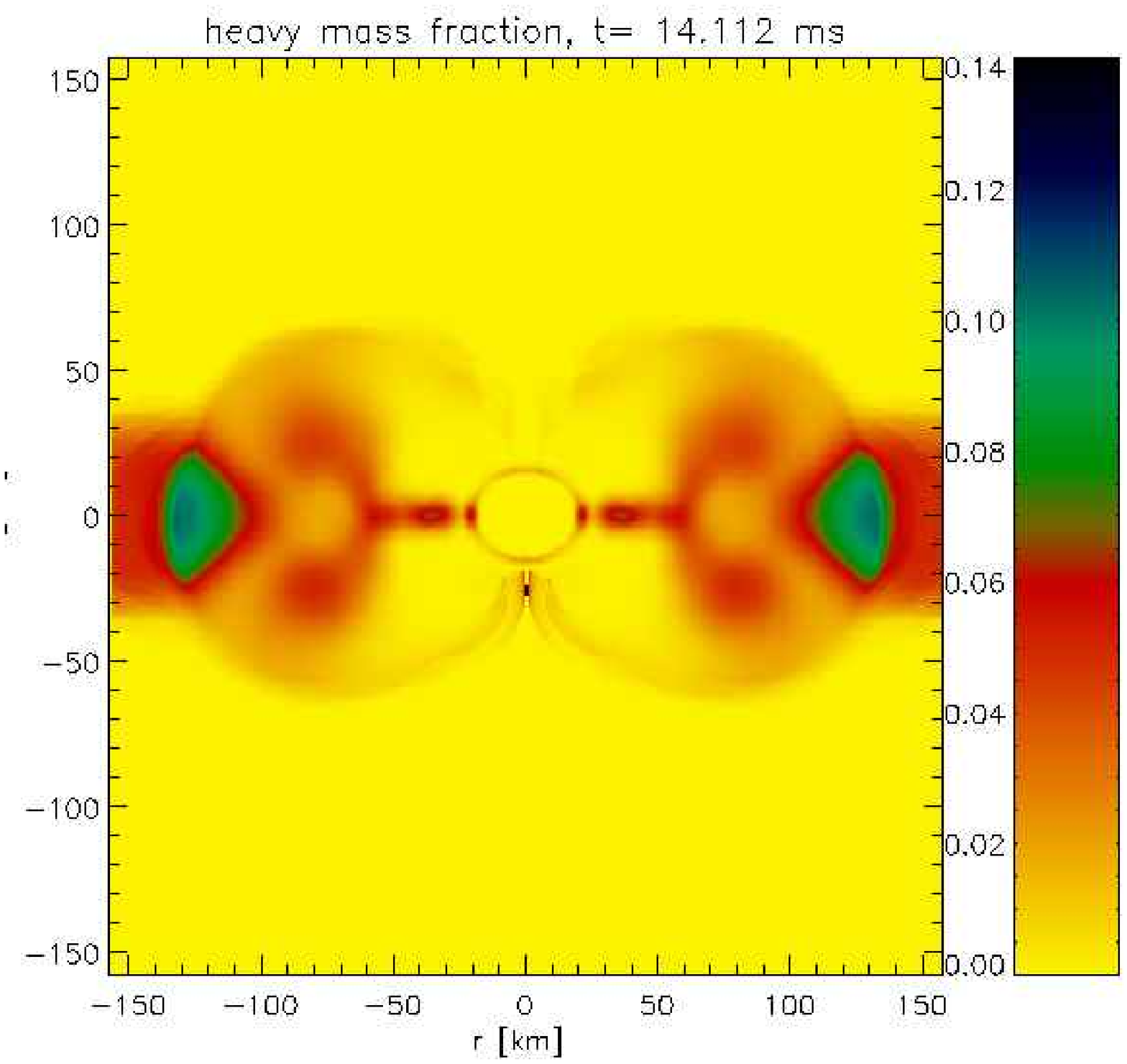,width=7cm,angle=0}}
\centerline{\psfig{file=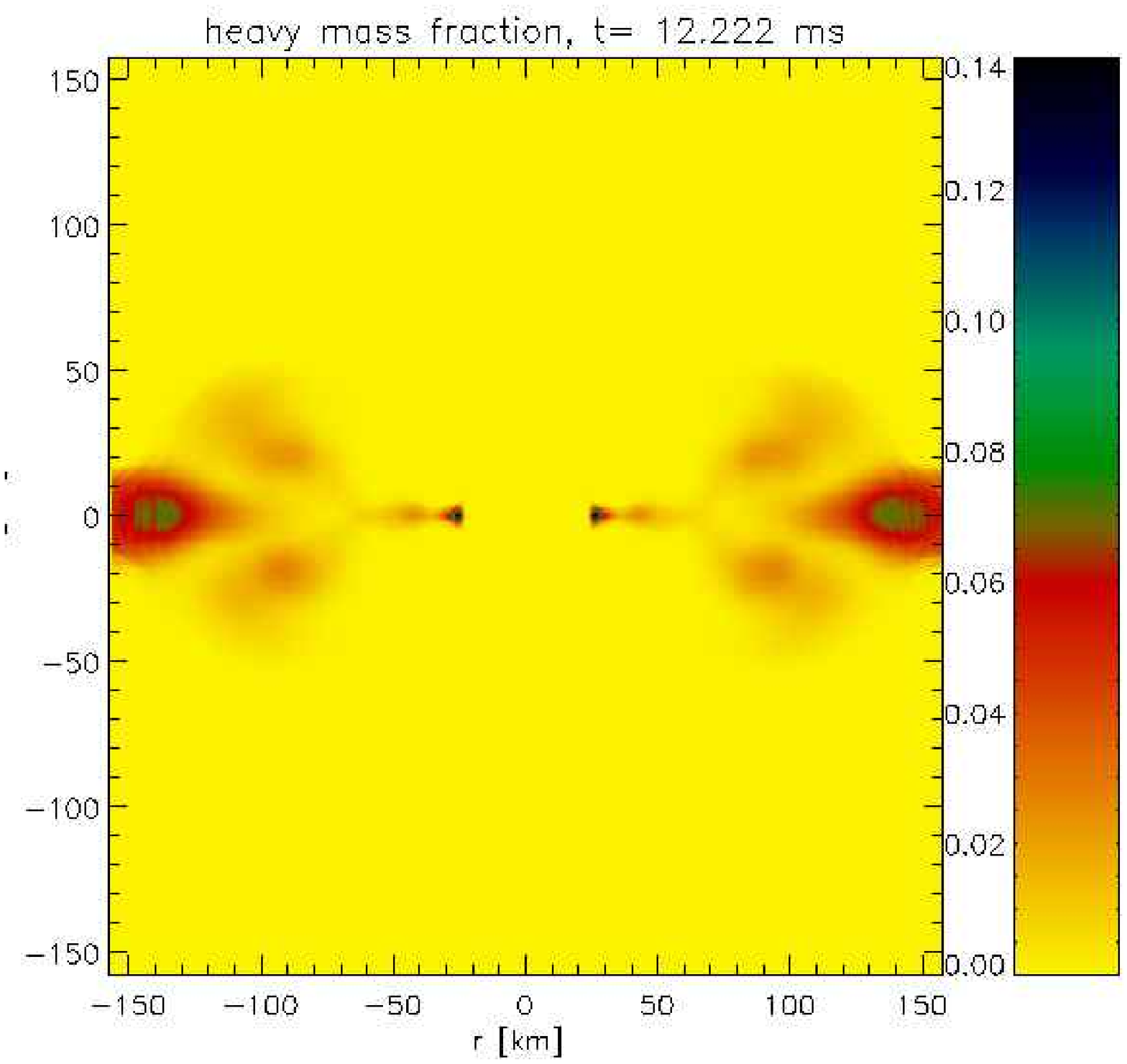,width=7cm,angle=0}}
\caption{\label{heavy_XZ}Shown are azimuthally averaged values of the heavy 
nucleus mass fraction for matter with $\rho > 10^{10}$ g cm$^{-3}$
(top to bottom run C, D and E). Although
heavy nuclei are present, neutrinos from the most luminous regions (see Fig.
\ref{em_geom}) can stream out vertically without encountering substantial
amounts of heavy nuclei. Therefore the latter ones do not influence the
total luminosity and the average energies.}
\end{figure}

Our scheme accounts for the coherent scattering of neutrinos off heavy nuclei,
for details we refer to the Appendix.
We had realized that, despite the high temperatures encountered in the disk,
matter finds it energetically favorable to form a non-negligible mass fraction
of heavy nuclei (paper I).
Due to the (approximate) proportionality of 
the scattering cross-section to the {\em square} of the nucleon number of the
heavy nucleus, see (\ref{coh_scat_formula}), nuclei could possibly dominate 
as an opacity source. To estimate how important the heavy nuclei really are 
for the neutrino emission,
we perform the following test. We take one time slice (t= 14.1 ms) of
our generic case, run C, update the  neutrino grid (see Appendix)
and then calculate with these opacities the properties of the emitted 
neutrinos. In one case we use --like in the dynamical simulation--
the full set of abundances given by the EOS
for both the emission and absorption/scattering processes and in the 
other case we assume the matter to be completely
dissociated into nucleons, i.e. the mass fractions are given by
\begin{equation}
x_p= Y_e, \quad x_n= 1-Y_e, \quad x_{\alpha}= 0, \quad x_h= 0.
\end{equation}
We find almost exactly the same numbers for both the mean energies and the
total luminosities, maximum deviations in the (more sensitive) total 
luminosities are below 5 \%.\\
The reason for this lies in the geometry of the heavy nucleus distribution.
In Fig. \ref{heavy_XZ} we show the azimuthally averaged values of the heavy 
nucleus mass fraction, $x_h$, of runs C, D and E;
these values are shown for matter with densities above 
10$^{10}$ gcm$^{-3}$, below that density matter is transparent to neutrinos.
Nuclei are present in the cool, equatorial inflow regions identified in 
paper I, see Figure 15 in Rosswog and Davies (2002). The butterfly-shaped 
temperature distribution is also reflected
in the nucleus mass fraction. It is interesting to note that despite the 
extreme temperatures in the central object a thin, nuclear crust can 
survive in our coolest case (run D). The hottest case (run E) is essentially 
free of heavy nuclei.
By comparing Fig. \ref{heavy_XZ} with Fig. \ref{em_geom}, right column, it 
becomes obvious that the neutrinos from the most luminous regimes can escape 
in each case vertically without having to pass through material containing 
an interesting amount of heavy nuclei. Therefore the influence of the heavy 
nuclei onto the total luminosity and the mean energies is negligible.

\subsection{Optical depths, neutrino-''spheres''}

\begin{figure}
\vspace*{-0.5cm}
\centerline{\psfig{file=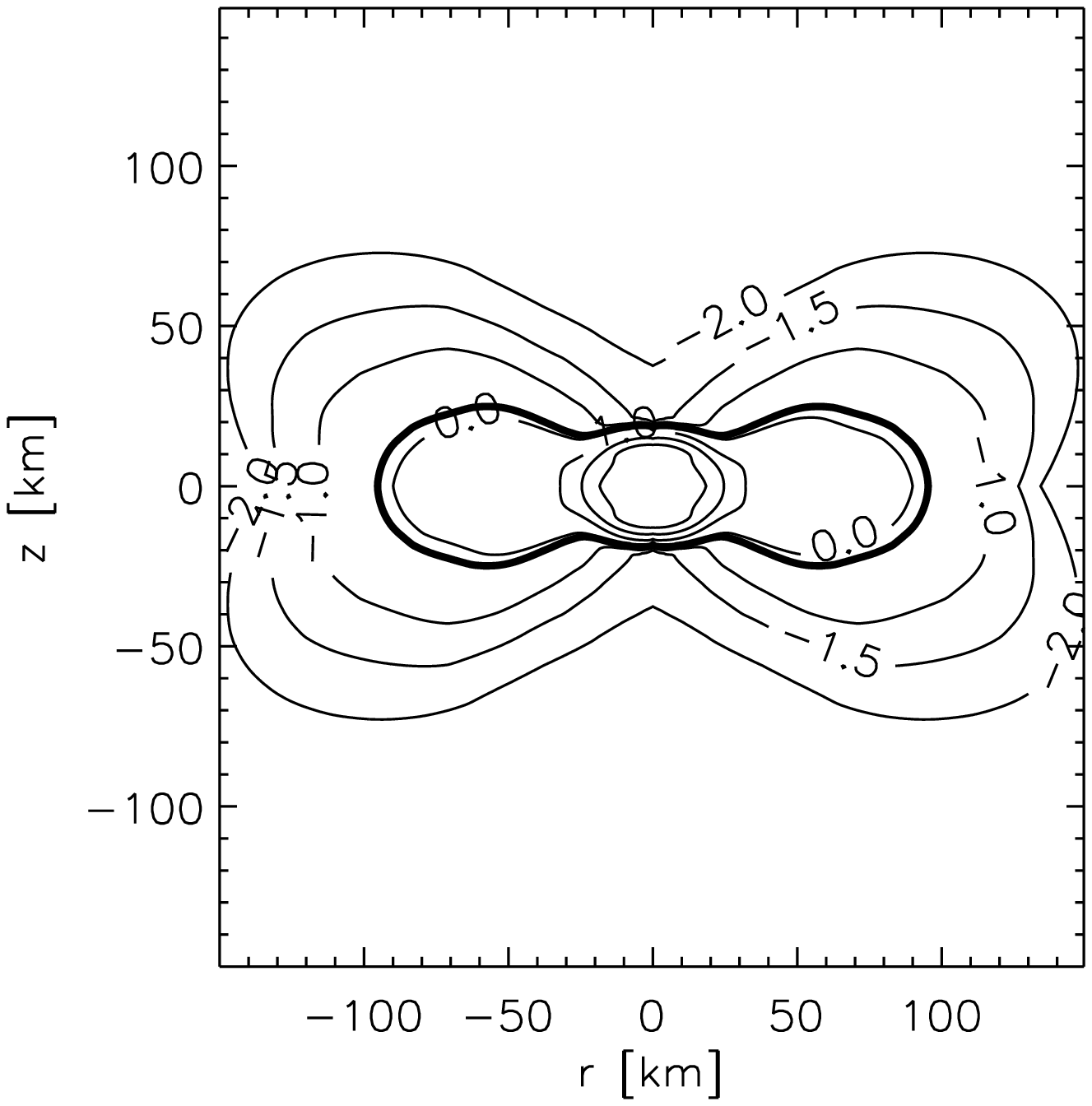,width=7.cm,angle=0}}
\vspace*{-0.5cm}
\centerline{\psfig{file=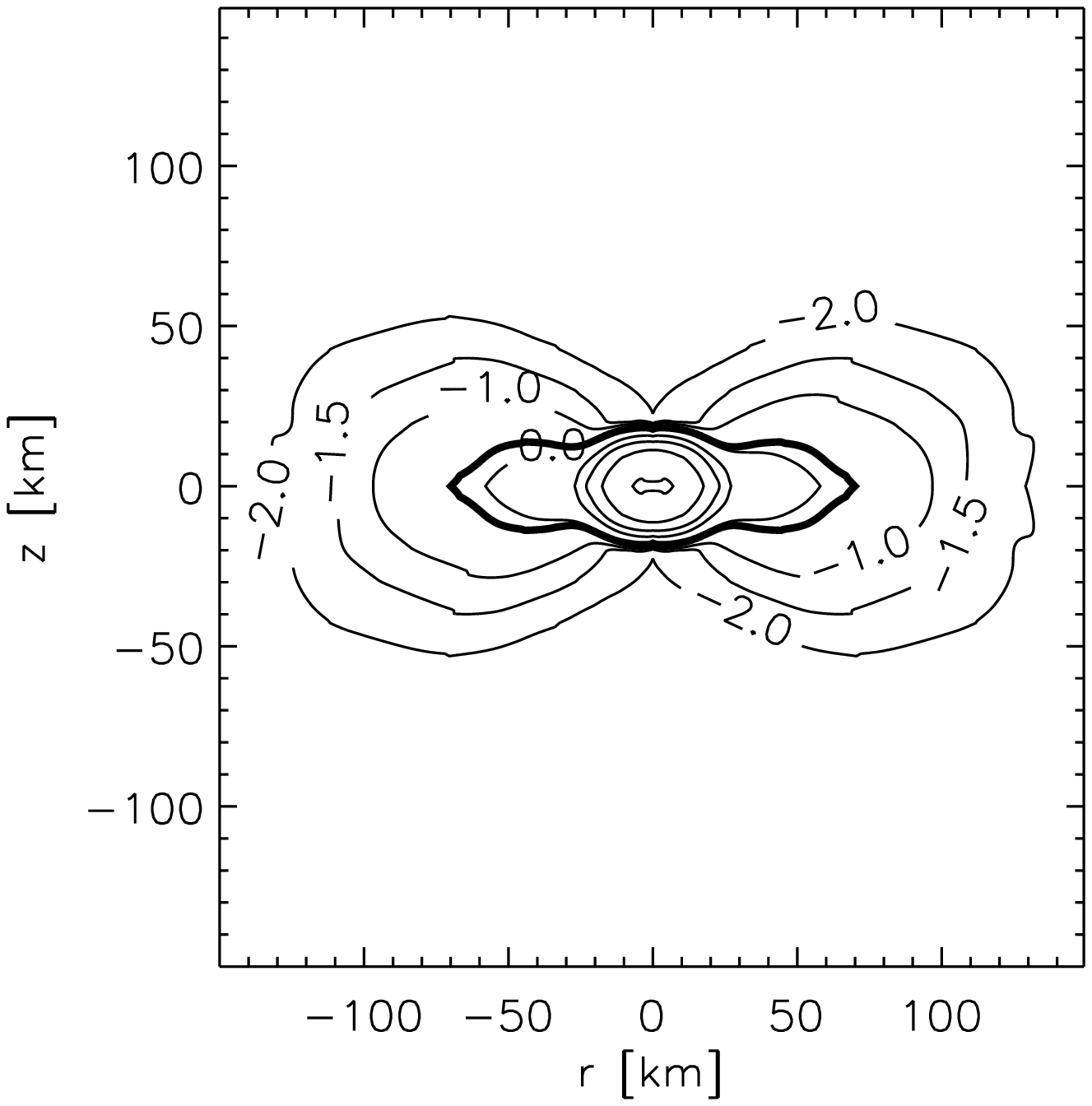,width=7.cm,angle=0}}
\vspace*{-0.5cm}
\centerline{\psfig{file=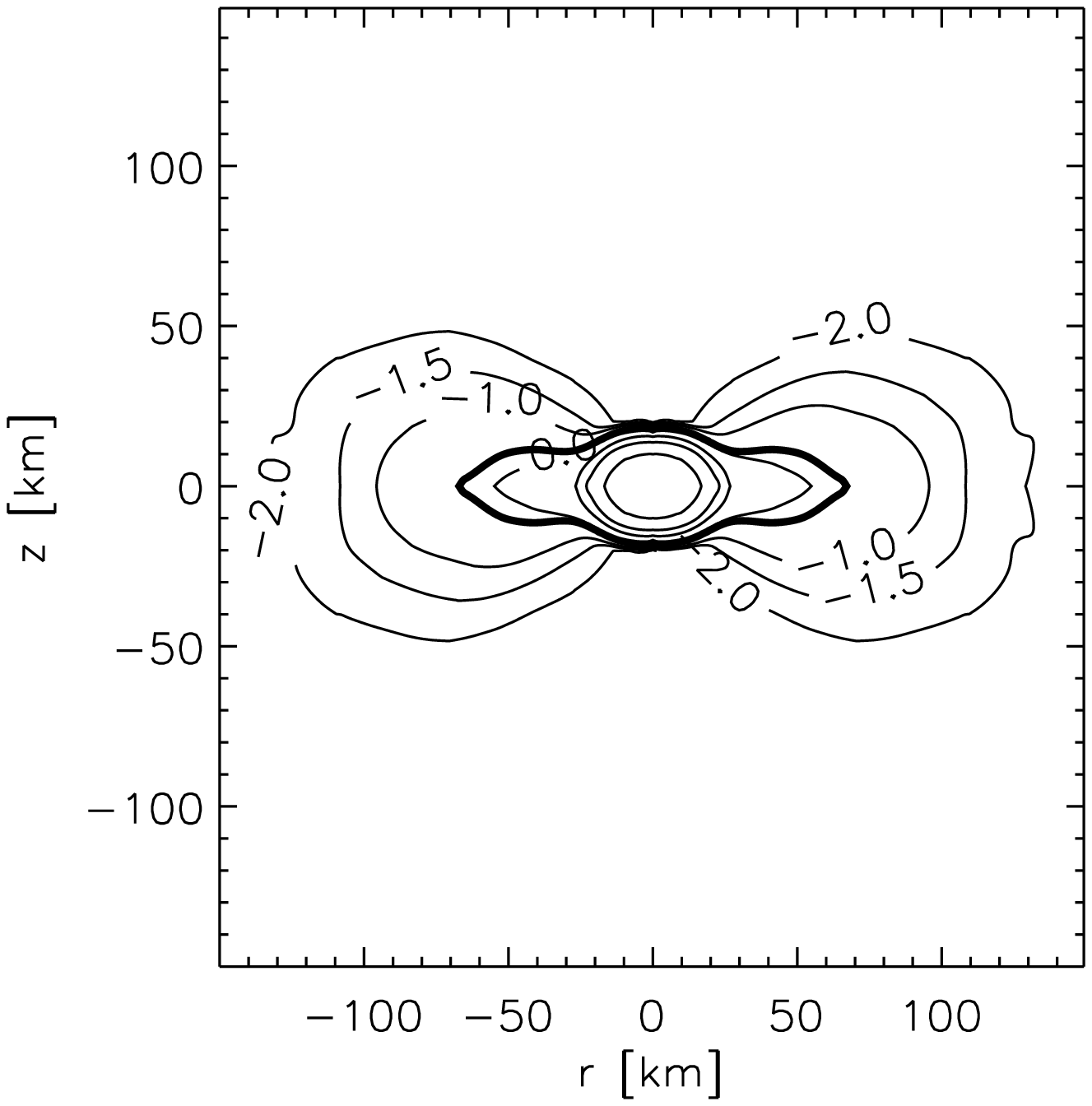,width=7.cm,angle=0}}
\caption{\label{tau_ir}End of run C (2 x 1.4 $\msun$, no spins):
logarithm of the optical depths (in steps of 0.5) of the various
neutrino flavours 
calculated on our grid. The first panel shows $\nu_e$, the second
$\bar{\nu}_e$ and the third $\nu_x$.
The thick lines give the locus of the ``neutrino sphere'' defined as
$\tau_{\nu_i}=$ 2/3, which is essentially the locus where the neutrinos
decouple from the debris matters. For the $\bar{\nu}_e$ and $\nu_x$ 
the neutrino 
spheres almost coincide since both are subject to scattering
processes, the absorption of $\bar{\nu}_e$ onto protons is only a minor 
correction. $\nu_e$ are additionally absorbed in the neutron-rich 
environment. Therefore the corresponding neutrino sphere is substantially 
larger.}
\end{figure}

\begin{figure}
\vspace*{-0.5cm}
\centerline{\psfig{file=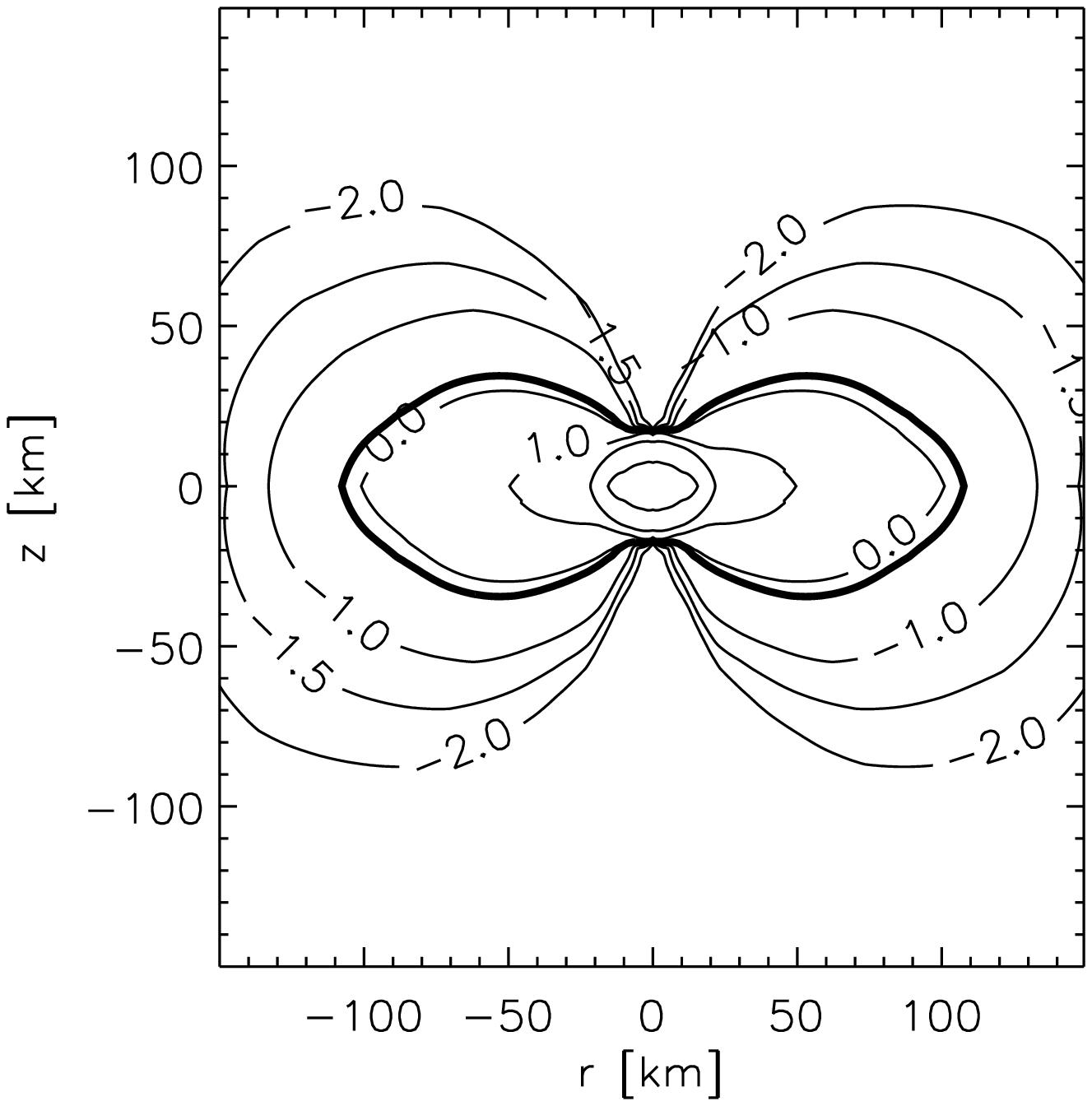,width=7.cm,angle=0}}
\vspace*{-0.5cm}
\centerline{\psfig{file=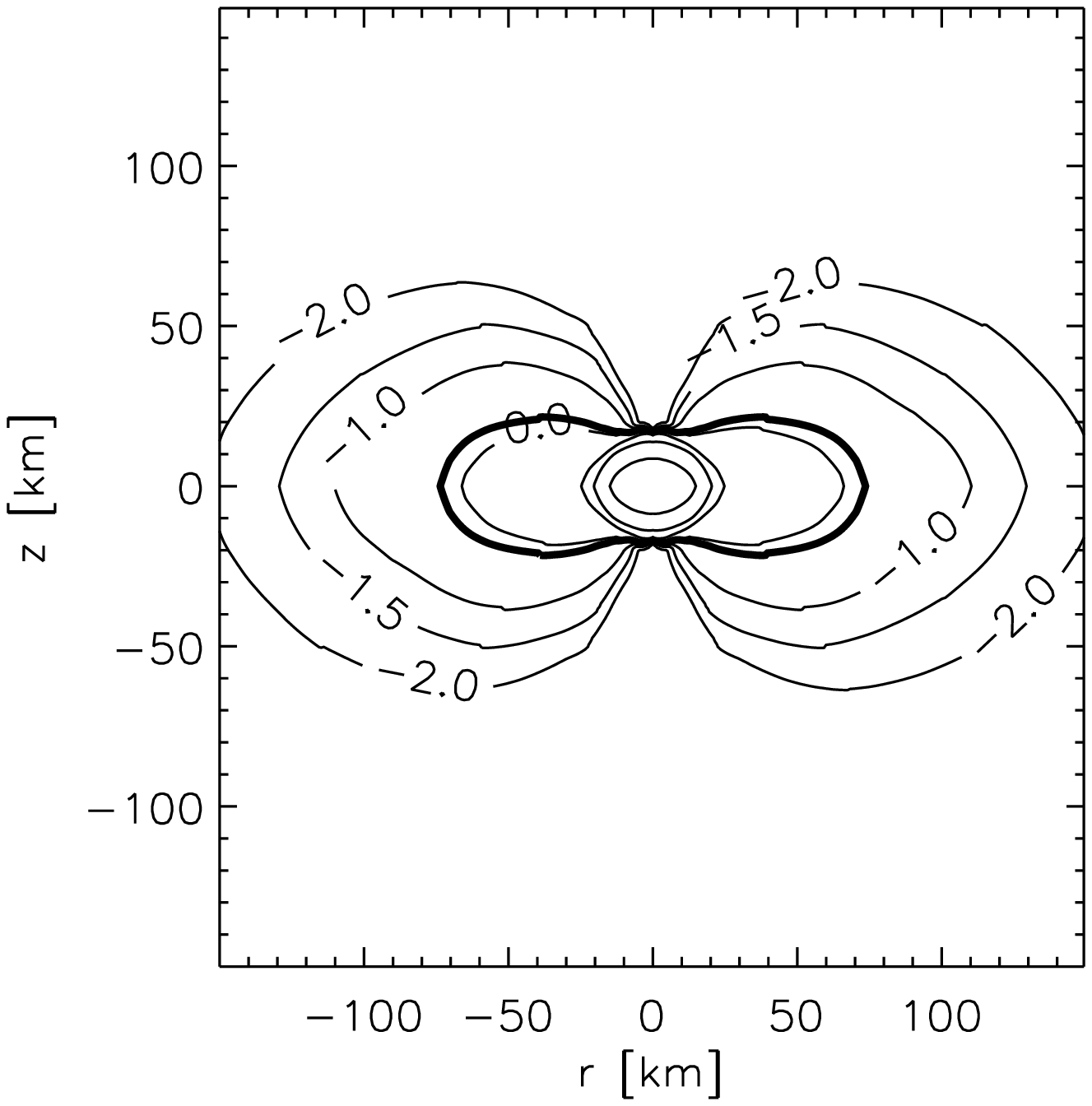,width=7.cm,angle=0}}
\vspace*{-0.5cm}
\centerline{\psfig{file=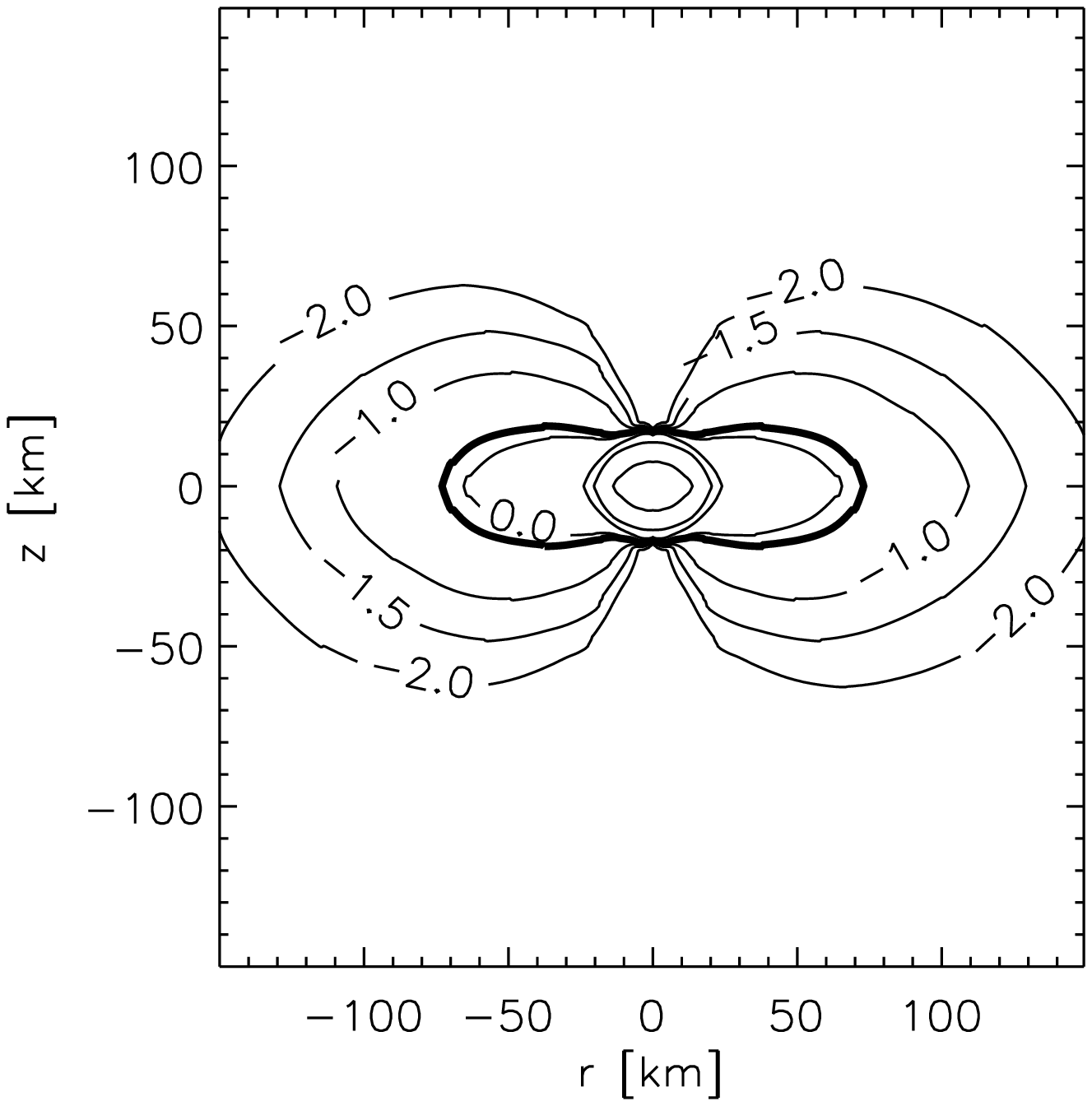,width=7.cm,angle=0}}
\caption{\label{tau_cor}Same as Figure \ref{tau_ir}, but for the
end of run D (corotation, 2 x 1.4 $\msun$).}
\end{figure}

\begin{figure}
\vspace*{-0.5cm}
\centerline{\psfig{file=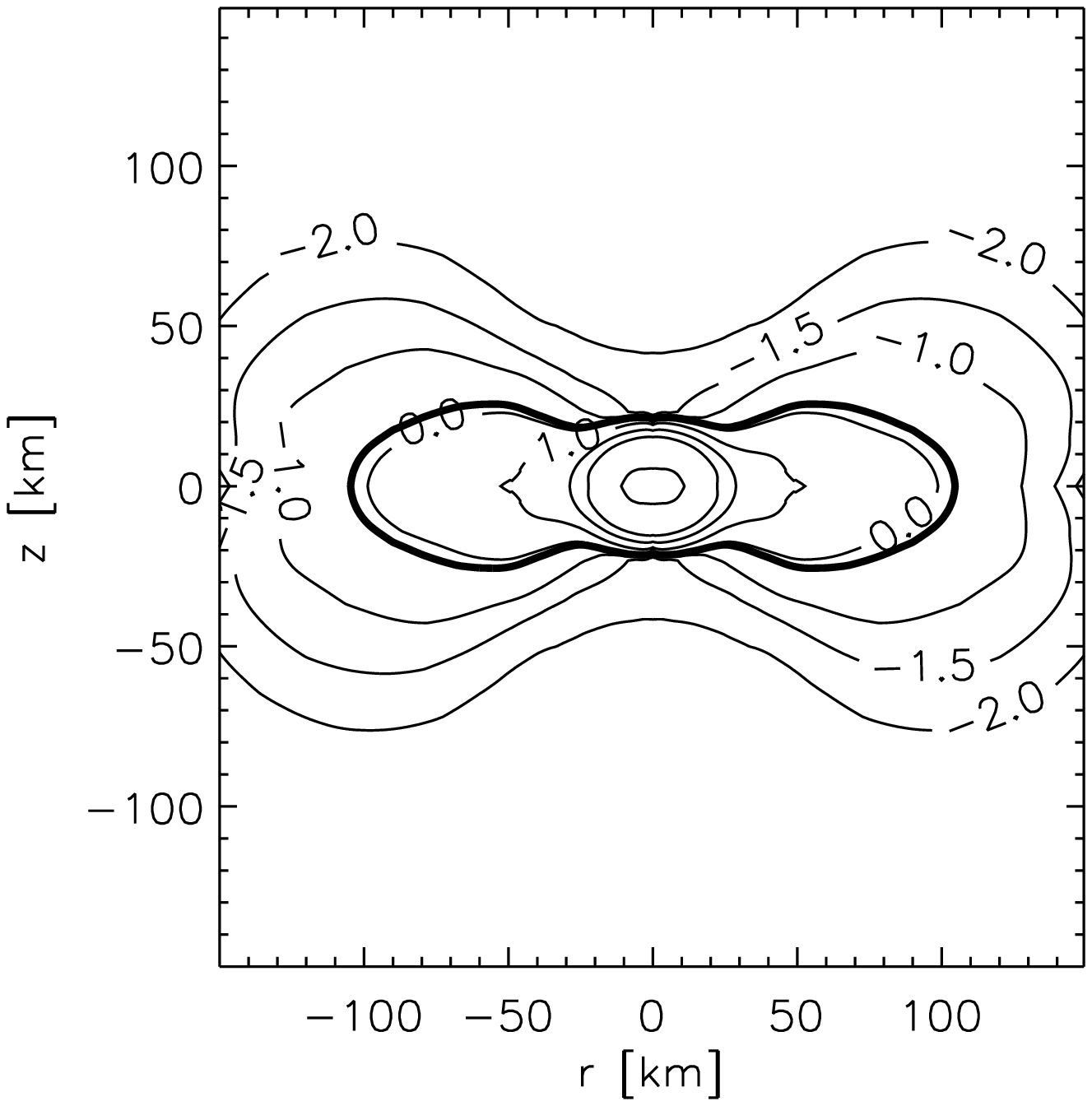,width=7cm,angle=0}}
\vspace*{-0.5cm}
\centerline{\psfig{file=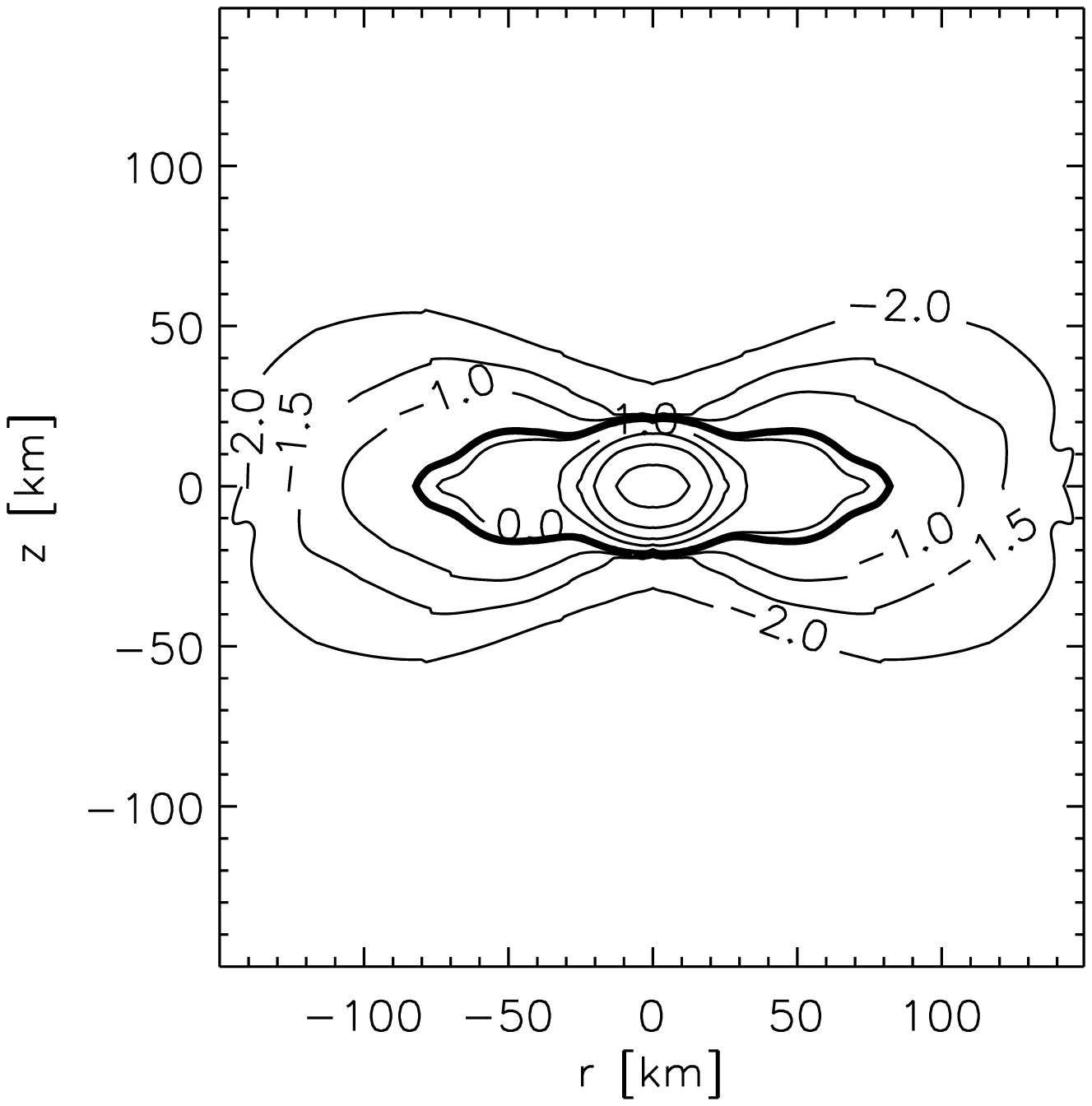,width=7cm,angle=0}}
\vspace*{-0.5cm}
\centerline{\psfig{file=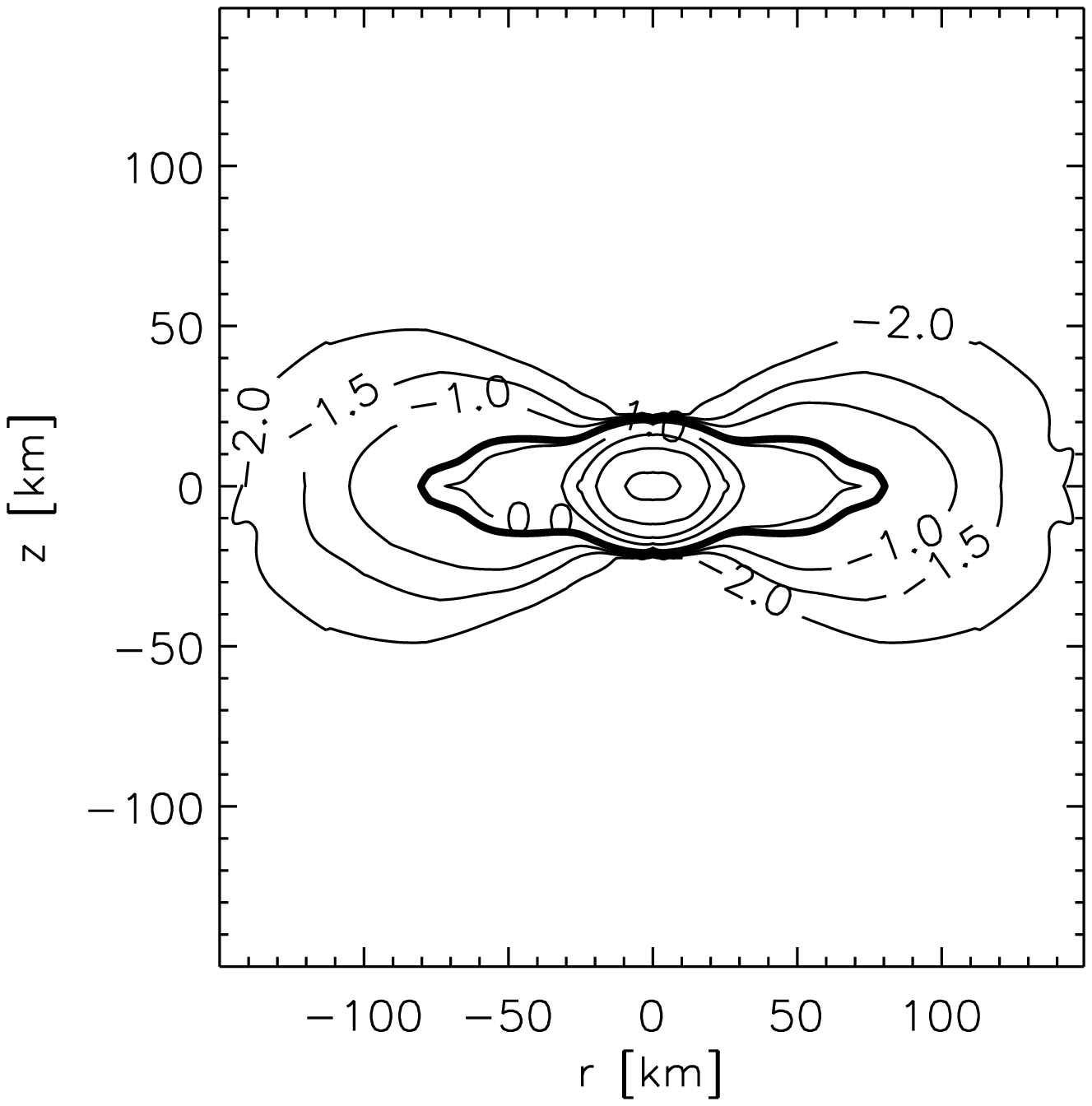,width=7cm,angle=0}}
\caption{\label{tau_irr2}Same as previous plot for run E.}
\end{figure}

To illustrate how the opaque matter is distributed in the merger remnant
we plot in Figures \ref{tau_ir} to \ref{tau_irr2}
contours of the spectrally averaged neutrino optical depth (see eq. 
(\ref{chi}) and (\ref{fermi_int})),
\begin{equation}
\tau_{\nu_i}= \chi_{\nu_i} \langle E^2_{\nu_i}\rangle = \chi_{\nu_i} 
\frac{F_4(\eta_{\nu_i})}{F_2(\eta_{\nu_i})} (k T)^2
\end{equation}
for all neutrino species at the end of run C, D and E. 
The $F_n$ are the standard Fermi integrals, see Appendix A.
Consistent with our neutrino treatment we have used the 
equilibrium values (see eq. (\ref{beta_eq})) for the degeneracy, $\eta$, 
of the $\nu_e$/$\bar{\nu}_e$ and have assumed a vanishing degeneracy parameter
 for the $\bar{\nu}_x$. 
In each of the Figures the uppermost panel shows contours of the optical depth
of the electron neutrinos, the middle panels refer to the electron 
anti-neutrinos and the remaining ones to the heavy lepton neutrinos.

\begin{figure*}
\psfig{file=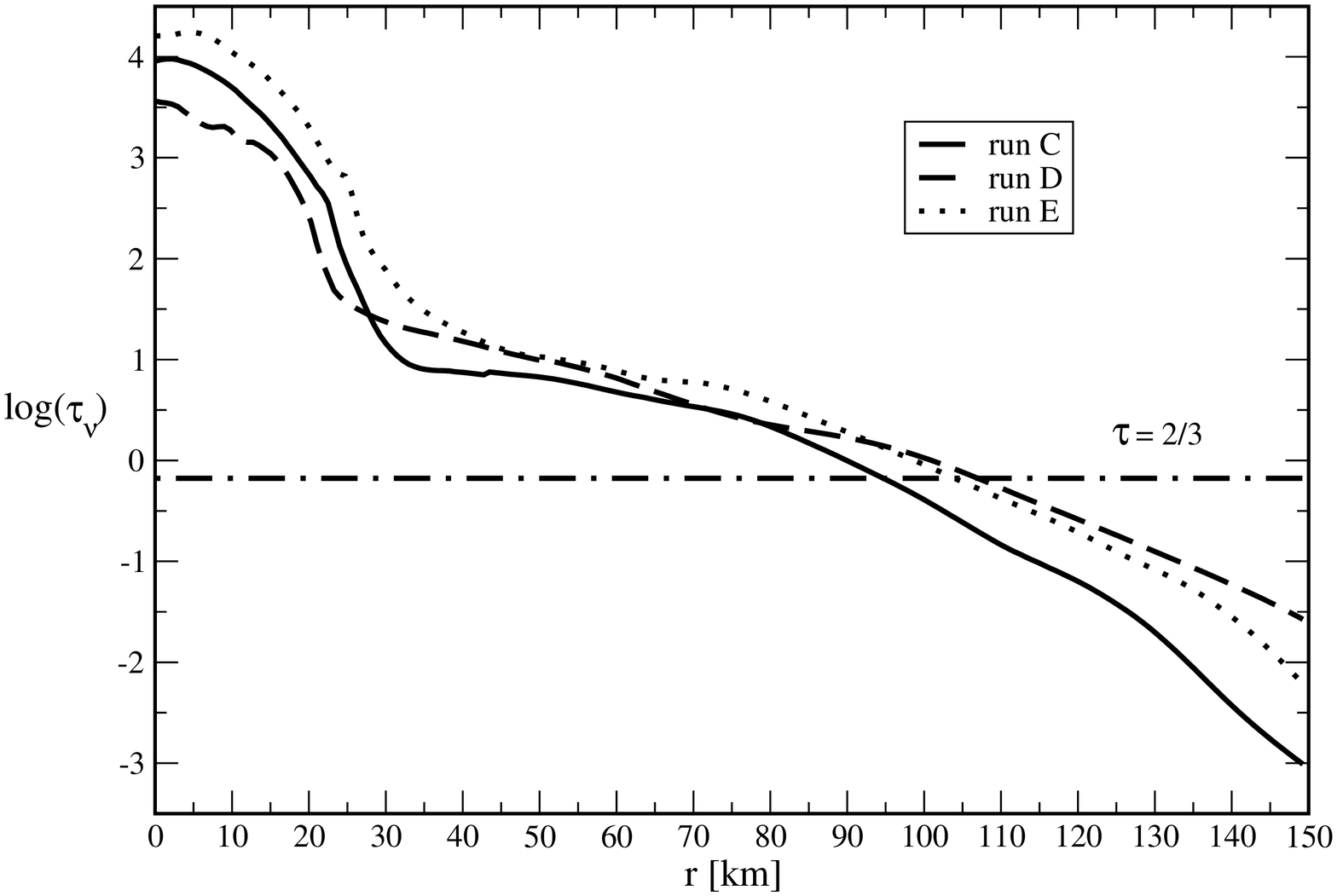,width=7cm,angle=-90}

\psfig{file=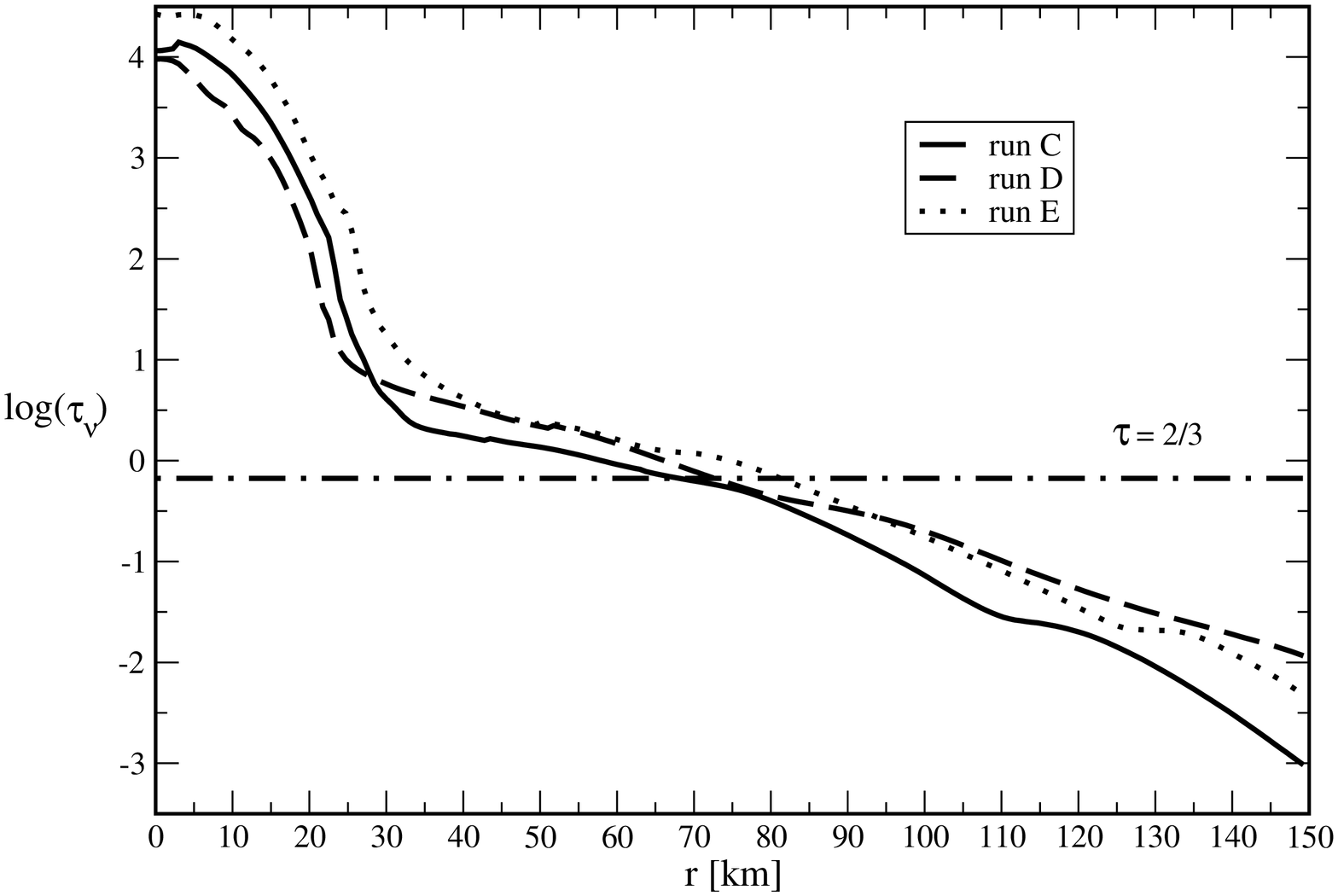,width=7cm,angle=-90}

\psfig{file=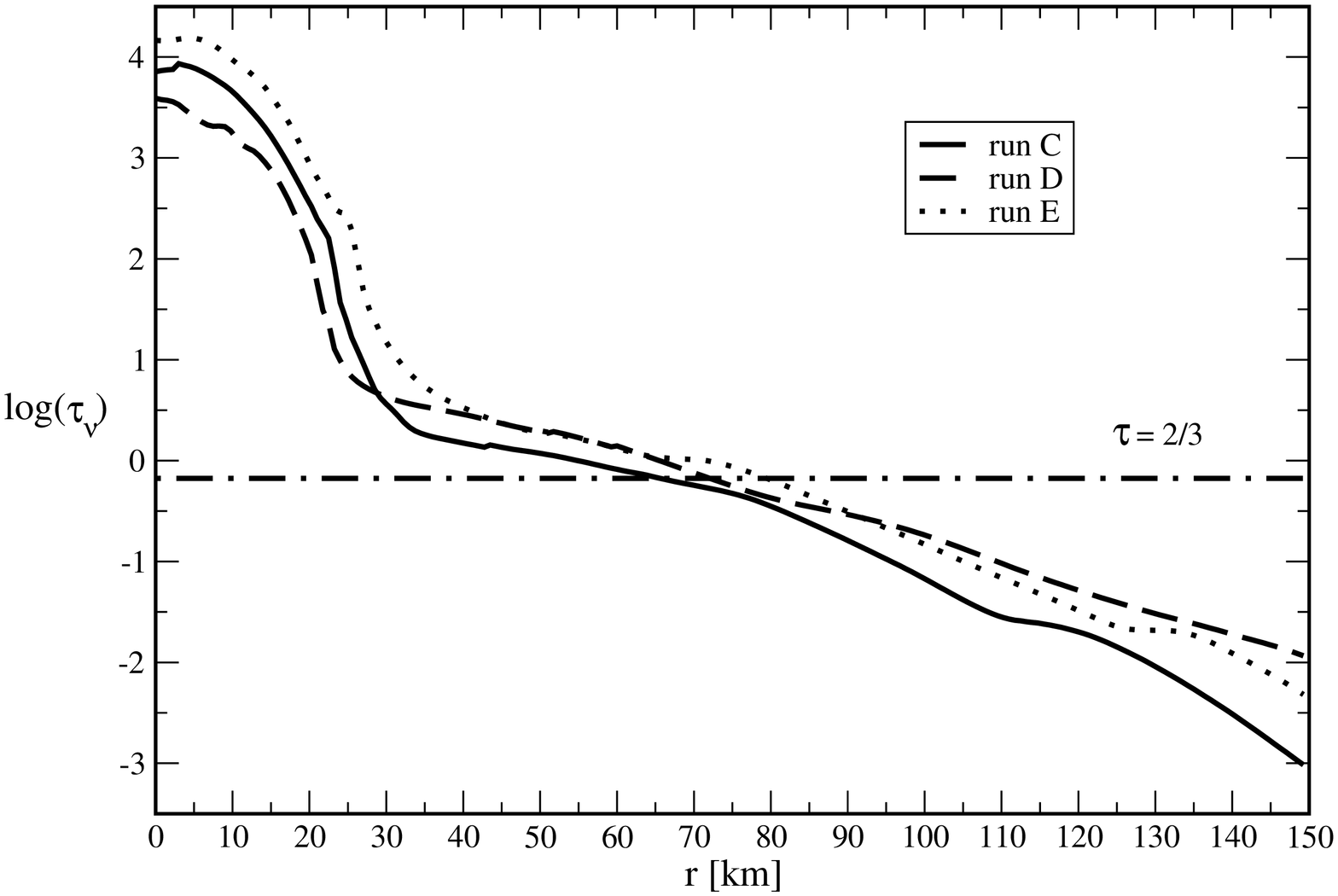,width=7cm,angle=-90}
\caption{\label{tau_z0} Optical depths at a height of z= 0 for the various
neutrino flavours (top to bottom: $\nu_e$, $\bar{\nu}_e$, $\nu_x$).}
\end{figure*}

The debris matter is most opaque to the electron-type neutrinos
which in addition to scattering are also absorbed onto the copiously available
free neutrons. Electron-type anti-neutrinos see matter less opaque and matter
is most transparent to the $\nu_x$.
We also show (as the  thick line) the ``neutrino-sphere'', defined 
as the locus with $\tau_{\nu_i}$= 2/3.
For the $\bar{\nu}_e$ and the $\nu_x$ the neutrino-spheres almost
coincide, with radial extensions of $\sim 70$ km and peak heights of
$\sim 20$ km, since both neutrino
types suffer essentially the same interactions (scattering events;
in the extremely neutron-rich debris absorption onto free protons is only 
a minor correction).
Due to their additional opacity sources the $\nu_e$ decouple
substantially further out, at radial extensions of $\sim 105$ km with 
peak heights of $\sim 35$ km. \\
The optical depths at height z= 0, i.e. in the orbital plane, are shown
in Figure \ref{tau_z0} (from top to bottom: $\nu_e$, $\bar{\nu}_e$ and 
$\nu_x$). In the central object values up to several times $10^4$ are 
reached, beyond $\sim 130$ km matter is essentially transparent to 
neutrinos of all types, i.e. $\tau_{\nu_i} < 0.1$.\\
It is interesting to note that it is only in the central object ($\tau_{\nu_i} 
>10$) that neutrinos are really trapped, see Figures \ref{tau_ir} to 
\ref{tau_z0}. At the edge of the central object, at distances of $\sim 30$ km
from the origin, the optical depth drops rapidly, but then only decreases
very slowly throughout the disk ($\sim 30$  km to  $\sim 100$ km). The whole
hot torus-region is therefore in the semi-transparent regime.

\subsection{Directional dependence of neutrino emission}
%
%

\begin{figure*}
\psfig{file=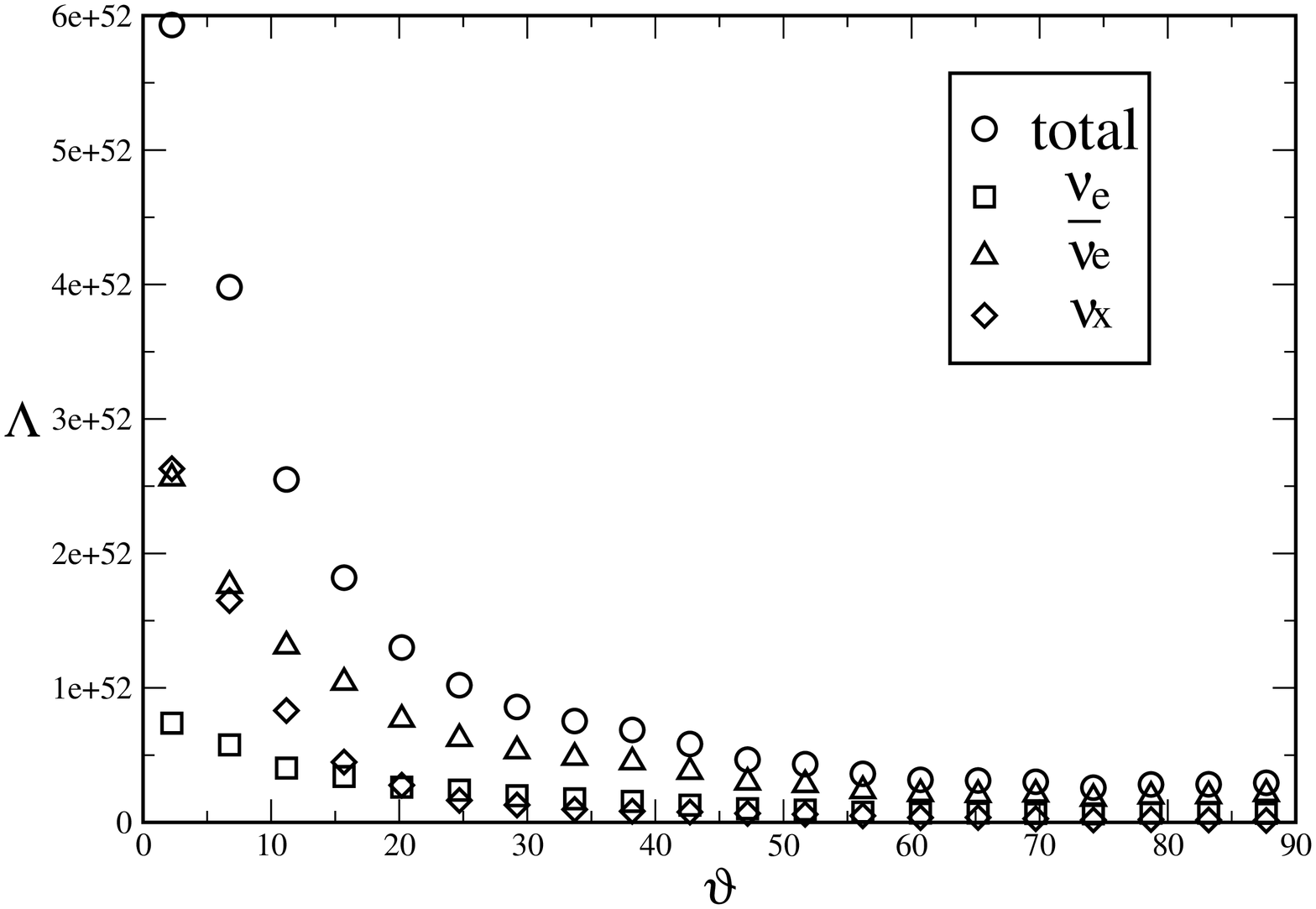,width=11cm,angle=-90}\\
\psfig{file=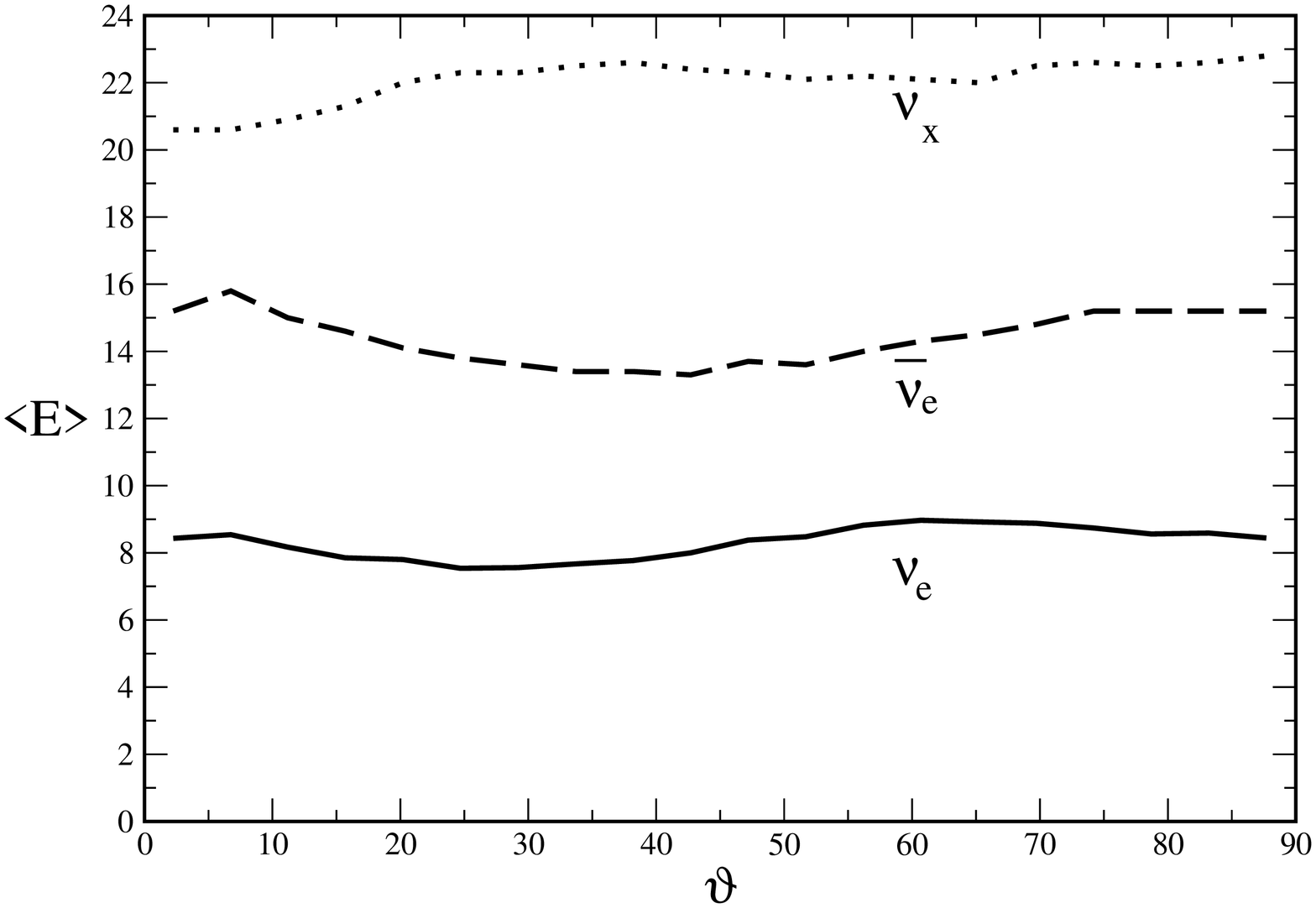,width=11cm,angle=-90}
\caption{\label{lambda_ir200}Directional dependence of the neutrino emission:
the upper panel shows the neutrino luminosity per solid angle, 
$\Lambda_{\nu_i}(\vartheta)= \frac{\Delta L_{\nu_i}}{\Delta \Omega}$ as a 
function of the angle (with the original binary rotation axis) under which
the system is observed. The apparent luminosity is given by 
$L^{\rm app}_{\nu_i}=
4 \pi \Lambda_{\nu_i}$. The lower panel shows the rms average energies of
the neutrinos as a function of $\vartheta$.}

\end{figure*}

It is consistent with our approach from eqs. (\ref{Ref}) and (\ref{Qef})
to think of the neutrinos emitted from an SPH-particle to be composed of
``free neutrinos'' and ``diffusive neutrinos''. In analogy to general 
diffusion equations we assume that the diffusive neutrino component is emitted 
in the direction of the local, negative density gradient, 
$\hat{n}= -(\nabla{\rho})/|-(\nabla{\rho})|$. 
We use the SPH-prescription to determine this density gradient at the position of
particle $i$:
\begin{equation}
\nabla{\rho}_i= \sum_j m_j \nabla_i W_{ij},
\end{equation}
where $m_j$ is the particle mass, 
$W_{ij}= W(\frac{|\vec{x}_i-\vec{x}_j|}{h_{ij}})$ the standard SPH-kernel 
(e.g. Monaghan 1992) and $h_{ij}$ is the arithmetic mean of the involved 
smoothing lengths. 
The free component, in contrast,
will emit isotropically. The fraction with which the both components contribute
to the neutrino luminosity of particle $j$ is given by
\begin{equation}
f_{\nu_i,Q_j}^{dif}= \frac{Q^{ef}_{\nu_i,j}}{Q^{dif}_{\nu_i,j}}
\quad {\rm and} \quad 
f_{\nu_i,Q_j}^{loc}= \frac{Q^{ef}_{\nu_i,j}}{Q^{loc}_{\nu_i,j}}.
\end{equation}
It can be easily checked that $f_{\nu_i,Q_j}^{dif}+  f_{\nu_i,Q_j}^{loc}=1$ by using
eqs. (\ref{Ref}) and (\ref{Qef}) and that the fractions approach their obvious limits
in the high and low-density regimes. Similarly, fractions of the emitted neutrino number,
 $f_{\nu_i,R_j}^{dif}$ and $f_{\nu_i,R_j}^{loc}$ can be defined as above, but with 
number emission rates per volume, $R_{\nu_i,j}$, rather than with energy emission rates 
per volume, $Q_{\nu_i,j}$. With these definitions the neutrino luminosity per solid angle, 
composed of a diffusive and a free component, is determined by 
\begin{equation}
\Lambda_{\nu_i}(\vartheta)= \frac{\Delta L_{\nu_i}}{\Delta \Omega}
=\frac{\sum_k \tilde{Q}_{\nu_i,k}^{dif}(\vartheta)}{2 \pi \sin (\vartheta) \Delta \vartheta} + 
\frac{\sum_j \tilde{Q}_{\nu_i,j}^{loc}}{4\pi} 
\label{Lambda}.
\end{equation}
Here $\tilde{Q}_{\nu_i,k}$ is the  $\nu_i$ energy emission rate  of  particle $k$,
from either ``diffusive'' neutrinos, $\tilde{Q}^{dif}_{\nu_i,k}= f_{\nu_i,Q_k}^{dif} 
\cdot \tilde{Q}^{ef}_{\nu_i,k}$  or ``free'' neutrinos,  $\tilde{Q}^{loc}_{\nu_i,k}= 
f_{\nu_i,Q_k}^{loc} \cdot \tilde{Q}^{ef}_{\nu_i,k}$.
In the above equation the $j$-sum extends over all the particles (since their ``free'' neutrinos 
radiate isotropically), the $k$-sum, however, only extends over those particles that
radiate into a ring of width $\Delta \vartheta$ in the $\vartheta$-direction, 
$\vartheta-\Delta \vartheta/2 < \vartheta_k < \vartheta+\Delta \vartheta/2$, where $\vartheta_k$
is given by $\cos(\vartheta_k)=\hat{n}_k \cdot \hat{e}_z$.
An observer that sees the merger from an angle $\vartheta$ with respect to the
initial binary rotation axis (= z-axis) would thus infer 
an apparent luminosity of
\begin{equation}
L_{\nu_i}^{\rm app}(\vartheta)= 4 \pi \Lambda_{\nu_i}(\vartheta).
\end{equation}
The quantity $\Lambda_{\nu_i}(\vartheta)$ for our generic case, run C,
is shown in Fig. \ref{lambda_ir200} (the other runs yield similar 
results). The luminosity per solid angle is peaked towards the z-axis: 
a system observed ``pole-on'' ($\vartheta 
\approx 0^\circ$) will yield a total neutrino energy flux, given by 
$f_{\nu_i}(\vartheta)= \frac{\Lambda_{\nu_i}(\vartheta)}{R^2}$, $R$ 
being the distance to the source, that is around 20 times larger than 
that of a system that is observed ``edge-on'' ($\vartheta \approx 90^\circ$).
This preferential emission is visible for all neutrino flavours, but most 
pronounced in the case of the heavy lepton neutrinos, $\nu_x$.
The latter ones are produced in the enormously temperature dependent
reactions (\ref{pair}) and (\ref{plasma}) and therefore emerge from 
the hottest parts of the remnant, i.e. they are generated within or 
close to the flattened central object,  where the density gradients point
along the z-direction.

To infer the $\vartheta$-dependence of the average neutrino energies we use the quantitity
\begin{equation}
\epsilon_{\nu_i}(\vartheta)=\sqrt{\frac{\sum_j \tilde{R}^{ef}_{\nu_i,j} E^2_{\nu_i,j}
(f_{\nu_i,R_j}^{dif}\cdot \Theta_j(\vartheta) + f_{\nu_i,R_j}^{loc})
}{\sum_j \tilde{R}^{ef}_{\nu_i,j}
(f_{\nu_i,R_j}^{dif}\cdot \Theta_j(\vartheta) + f_{\nu_i,R_j}^{loc})}}  .
\end{equation}
Here, the $j$-sum extends over all the particles; to count only the contributing particles
in the diffusive part, the function 
\[
\Theta_j(\vartheta)= \left\{ \begin{array}{r@{ }l}
                         &2/(\sin(\vartheta) \Delta \vartheta)  \quad \mbox{for } \quad \vartheta-\Delta \vartheta/2 < \vartheta_j < \vartheta+\Delta \vartheta/2\nonumber\\
                         &0  \;\quad\quad\quad\quad\quad\quad \mbox{else}\nonumber         \end{array}
                             \right.  \]
has been introduced.
The distribution of the average energies, see second panel Fig. \ref{lambda_ir200}, 
is relatively flat for all neutrino types.

\section{Summary and Discussion}

We have presented the neutrino emission results from our high-resolution 
simulations of the coalescence of two neutron stars. We find typically 
total neutrino luminosities of $\sim 2 \cdot 10^{53}$ erg/s with rms energies
of $\sim 8$ MeV for electron type neutrinos, $\sim 15$ MeV for their 
anti-particles and $\sim 20 - 25$ MeV for the $\mu$- and $\tau$-neutrinos and 
their antiparticles. We have performed two runs that are intended to give
an upper and a lower limit to the total neutrino luminosities: in the case of
initial corotation the stars merge extremely smoothly. This goes along with 
moderately high matter temperatures and neutrino luminosities lower by a 
factor of two than in our standard case. In the other extreme case we consider
the coalescence of two 2.0 $\msun$ neutron stars without initial spins. This 
case leads to the hottest merger remnant and, correspondingly, to the highest 
neutrino luminosities, around $4 \cdot 10^{53}$ erg/s.\\
The contributions of the extremely hot, but also neutrino-opaque central 
objects are marginal, typically only a few percent. Most neutrinos are
produced in the debris torus around the central object, which exhibits 
temperatures well above the positron production threshold and which is 
very neutron rich ($Y_e \sim 0.1$). These conditions favour positron 
captures on free neutrons over electron captures and therefore yield
neutrino luminosities which are clearly dominated by the $\bar{\nu}_e$.
The heavy lepton neutrinos contribute only 
$\sim 10 \%$ to the total luminosity. The $\nu_e$ and $\bar{\nu}_e$
are predominantly produced in electron and positron captures on free nucleons,
only $\sim 10\%$ come from the pair and plasma process. The whole disk, 
with distances of $\sim30$ km to $\sim$ 100 km is semi-tranparent to
the neutrinos; it is only within the high-density central object that they
are really trapped ($\tau_{\nu_i} >10$).\\
We find qualitative agreement with the results described in Ruffert 
and Janka (2001) as far as the dominance of the $\bar{\nu}_e$ and the 
hierarchies in the rms neutrino energies 
($\epsilon_{\nu_e} < \epsilon_{\bar{\nu}_e} < \epsilon_{\nu_x}$) are
concerned. Our total luminosities and rms energies, however, are  
lower than those found in their models, typically by a factor of $\sim 2$
in the luminosities and around $\sim 20 \%$ in the mean energies.
This comes in part from the fact that not exactly the same initial conditions 
are used: the 'standard' mass they use is 1.6 $\msun$ rather than our value, 
1.4 $\msun$. This mass difference is 
expected to lead to slightly increased luminosities. Another difference 
is the EOS. The EOS of Shen et al. (1998) that we use is in the density
regime of 12 $<$ log($\rho$) $<$ 14, where large fractions of the 
neutrino emission stems from, substantially stiffer than the 
Lattimer-Swesty EOS that Ruffert \& Janka use (compare Fig. 2 in Rosswog \&
Davies 2002). This leads to a less compact configuration with lower 
temperatures and correspondingly lower neutrino luminosities in our case. 
Further possibilities include a different amount of numerical viscosity 
(see Rosswog \&  Davies 2002 for a discussion) in both 
codes and maybe the interaction with the background medium in the simulations 
of Ruffert \& Janka (2001; see their paper for a discussion of this point).
Finally, the lower luminosities may also come from differences in the 
leakage prescriptions. However, as shown in the appendix, if at all, our 
scheme tends to {\em overestimate} the luminosities. Therefore the true 
neutrino luminosities could be even lower, a fact that has serious 
consequences for the ability of neutron star mergers to produce a 
gamma-ray burst fireball via neutrino annihilation. This is discussed
further in Rosswog \& Ramirez-Ruiz (2002) and Rosswog et al. (2003).\\
Despite the high temperatures we find areas in the disk that contain a
substantial mass fraction of heavy nuclei. One might expect this to influence
the neutrino luminosities, since the coherent scattering cross sections are
$\propto A^2$, where $A$ is the nucleon number of the nucleus. This, however,
is not the case. In each of the investigated cases the most neutrino-luminous
parts of the remnant are essentially free of heavy nuclei and the neutrinos
can always escape via almost completely photo-dissociated matter.\\
We find that the neutrino emission per solid angle is focussed towards
the initial binary rotation axis. A merger remnant observed ``pole-on''
has an apparent neutrino luminosity that is about 20 times larger
than a remnant seen ``edge-on''.\\
A typical neutron star merger produces mean neutrino energies
very similar to those resulting from the core collapse of a massive star.
A distinctive signature between both events is the strong 
dominance of the electron anti-neutrinos over electron neutrinos in the 
merger case. The most unique proof, however, for
neutrinos coming from a neutron star coalescence rather than from a SN would
be the nearly coincident detection of a binary ``chirp''-signal in 
gravitational waves. The peak luminosity in neutrinos will be reached about
15 ms after the peak in the gravitational wave luminosity.

\vspace*{1cm}
{\bf Acknowledgements}\\
\vspace*{0.1cm}\\
It is a pleasure to thank E. Ramirez-Ruiz, R. Speith and the Leicester theory group
for fruitful discussions. This work has benefited from the excellent
support from the Leicester supercomputer team Stuart Poulton, 
Chris Rudge and Richard West.
Most of the computations reported here were performed using the UK
Astrophysical Fluids Facility (UKAFF).
Part of this work has been performed using the University of Leicester 
Mathematical Modelling Centre's supercomputer which was purchased through 
the EPSRC strategic equipment initiative.
This work was supported by a PPARC Rolling Grant for Theoretical Astrophysics.
S.R. gratefully acknowledges the support of PPARC by an Advanced Fellowship.
M. L. acknowledges support by the NSF under contract AST-9877130
at the University of Tennessee, Knoxville and the Oak Ridge
National Laboratory, managed by UT-Batelle, LLC, for the U.S.
Department of Energy under contract DE-AC05-00OR22725.

\begin{appendix}
\section{Neutrino treatment}
The rates that we use in the simulations are smooth interpolations
between diffusion and local production rates.
If we denote for a given neutrino species  $\nu_i$ the number
emission rates by $R_{\nu_i}$ per volume and energy emission rates per volume 
by $Q_{\nu_i}$, our prescription for the  {\em effective} rates reads
    
\begin{eqnarray} 
	R_{\nu_i}^{ef} = 
                        R_{\nu_i}^{loc} \left(1+
                        \frac{R_{\nu_i}^{loc}}{R_{\nu_i}^{dif}} \right)^{-1}
                        \label{Ref}\\
	Q_{\nu_i}^{ef} = 
                        Q_{\nu_i}^{loc} \left(1+
	                \frac{Q_{\nu_i}^{loc}}{Q_{\nu_i}^{dif}}\right)^{-1}.
                        \label{Qef}
\end{eqnarray}
This ansatz is similar to the one used in Ruffert et al. (1996).
Here the quantities with the superscript ``$loc$'' denote the locally produced 
rates of number and energy while the superscript ``$dif$'' refers to the
diffusion rates that are further specified below. 
In the transparent regime, where the diffusion time scale $T^{dif}_{\nu_{i}}$
is short, and therefore $R_{\nu_i}^{dif} \gg R_{\nu_i}^{loc}$ and 
$Q_{\nu_i}^{dif} \gg Q_{\nu_i}^{loc}$ 
all the locally produced neutrinos stream out freely. In the very opaque 
regime, where $T^{dif}_{\nu_{i}}$ is large, the neutrinos leak out on the 
diffusion time scale. Therefore both limits are treated correctly, the 
regime inbetween these limits is handled via interpolation.\\
The mean neutrino energy of each SPH-particle (particle index suppressed) 
is then found from
\begin{equation}
	E^{ef}_{\nu_i}= \frac{\sum_r Q^{ef}_{\nu_i,r}}{\sum_r 
	                 R^{ef}_{\nu_i,r}}, \label{Eeff}
\end{equation}
where $r$ labels all reactions producing neutrinos of type ${\nu_i}$.
Note, that these (mean) energies are used exclusively for book-keeping 
purposes, in all places where a dependence on neutrino energies occurs, we 
integrate cross-sections over a Fermi-distribution (see below).\\
To characterize the average neutrino energies of the total system we use 
rms energies given by
\begin{equation}
\epsilon_{\nu_i}= \sqrt{ \frac{\sum_j \tilde{R}^{ef}_{\nu_i,j} 
(E^{ef}_{\nu_i,j})^2}{\sum_j \tilde{R}^{ef}_{\nu_i,j}}},
\label{mean_E}
\end{equation}
where $j$ labels the SPH-particles and $\tilde{R}^{ef}_{\nu_i,j}$ is the rate
of neutrino number emission of particle $j$ (not to be confused with the rate
{\em per volume}, $R^{ef}_{\nu_i,j}$).\\
We have tested this scheme in spherical symmetry against
stationary state Boltzmann transport (Mezzacappa \& Messer 1999). 
To this end we determined the neutrino properties
for a frozen matter background. The background properties ($\rho,T$ and $Y_e$)
were either taken from neutron star merger (Rosswog \& Davies 2002) or core 
collapse supernova simulations (Liebend\"orfer et al. 2002). 
While the rms neutrino energies agree within 20 $\%$ the 
accuracy of the luminosities depends on the importance of the semi-transparent
regime where the interpolation (eqs. (\ref{Ref}) and (\ref{Qef})) is applied. 
In the worst case we found that our scheme overestimates the luminosities
by a factor 3-4.

\subsection{Free Emission Rates}\label{app_em_rate}
In the following we will neglect the electron mass and the nucleon mass 
difference, $Q= m_n - m_p = 1.2935$ MeV, in all the cross sections 
(Tubbs \& Schramm 1975). This is 
appropriate for our purposes and largely simplifies the involved rate 
expressions. 
We further assume the neutrino temperature to be identical to the local matter
temperature and, where necessary, we assume the neutrinos to follow a 
Fermi-distribution. The chemical potentials of the $\nu_x$ are generally
assumed to vanish, for $\nu_e$ and $\bar{\nu}_e$ we apply the equilibrium 
values
\begin{equation}
    \mu_{\nu_e}= - \mu_{\bar{\nu}_e} = \bar{\mu}_e - \hat{\mu} - Q,
	\label{beta_eq}
\end{equation}
wherever they occur in the sequel. Here $\bar{\mu}_e$ is the electron chemical 
potential (with rest mass) and $\hat{\mu}$ is the difference in the neutron 
and proton chemical potentials (without rest mass). 
Degeneracy parameters $\mu_i/T$ are denoted by $\eta_i$, temperatures are 
always in units of energies.\\
With these approximations and ignoring momentum transfer to the 
nucleon (Bruenn 1985) 
 the {\em electron capture} rate per volume reads 
\begin{equation} 
    R_{EC}= \beta \; \eta_{pn} T^5 F_4(\eta_e),\label{REC},
\end{equation}
with 
\begin{equation}
 \beta= \frac{\pi}{h^3 c^2} \frac{1+3 \alpha^2}{(m_e c^2)^2} 
    \sigma_0.
\end{equation}
Here $h$ is Planck's constant and $c$ the speed of light, $\alpha 
\approx 1.25$, $m_e$ is the electron mass, $\sigma_0\approx 1.76\cdot 10^{-44}$cm$^2$. $F_n$ is a Fermi integral given by
\begin{equation} 
    F_n(z)= \int_0^\infty \frac{x^n dx}{e^{x-z}+1} \label{fermi_int}
\end{equation}
and can be efficiently evaluated via series expansions 
(Takahashi et al. 1978).
The factor $\eta_{pn}$ given by
\begin{equation} 
    \eta_{pn}= \frac{n_n-n_p}{exp(\hat{\mu}/T)-1}, 
\end{equation}
 takes into account the nucleon final state blocking and
reduces in the non-degenerate limit to the proton number density $n_p$, 
$n_n$ refers to the neutron number density.
Following the analogous procedure one finds for the energy emission rate
\begin{equation} 
    Q_{EC}= \beta \, \eta_{pn} T^6 F_5(\eta_e),
\end{equation}
and for the mean energy of the emitted neutrinos
\begin{equation} 
    \langle E_{\nu_e} \rangle_{EC}= \frac{Q_{EC}}{R_{EC}} = 
    T \frac{F_5(\eta_e)}{F_4(\eta_e)}.
\end{equation}	
The corresponding rates for {\em positron captures} read
\begin{equation} 
    R_{PC}= \beta \, \eta_{np} T^5 F_4(-\eta_e),
\end{equation}
\begin{equation} 
    Q_{PC}= \beta \, \eta_{np} T^6 F_5(-\eta_e),
\end{equation}
\begin{equation} 
    \langle E_{\bar{\nu}_e} \rangle_{PC}= 
    T \frac{F_5(-\eta_e)}{F_4(-\eta_e)},
\end{equation}
where $\eta_{np}$ is obtained from $\eta_{pn}$ by interchanging the neutron
and proton properties. \\
The ``thermal'' processes are taken into account via fit formulae. For the 
energy emission from the {\em pair process} we use the prescription of 
Itoh et al. (1996).
 The number emission rate is obtained by deviding by the mean energy per neutrino pair (Cooperstein et al. 1986)
\begin{equation} 
    \langle E_{\nu_i \bar{\nu}_i} \rangle_{pair}= 
    T \left( \frac{F_4(\eta_e)}{F_3(\eta_e)} 
    + \frac{F_4(-\eta_e)}{F_3(-\eta_e)} \right).
\end{equation}
For the {\em plasmon decay} we use the formulae of Haft et al. (1994)
with 
\begin{equation} 
    \langle E_{\nu_i \bar{\nu}_i} \rangle_{\gamma}= 
    T \left( 2 + \frac{\gamma^2}{1+\gamma} \right),
\end{equation}
where $\gamma= \gamma_0  \sqrt{\pi^2/3 + \eta_e^2}$ and $\gamma_0= 5.565
\cdot10^{-2}$.

\subsection{Diffusive Emission Rates}
\begin{figure}
    \hspace*{0cm}\psfig{file=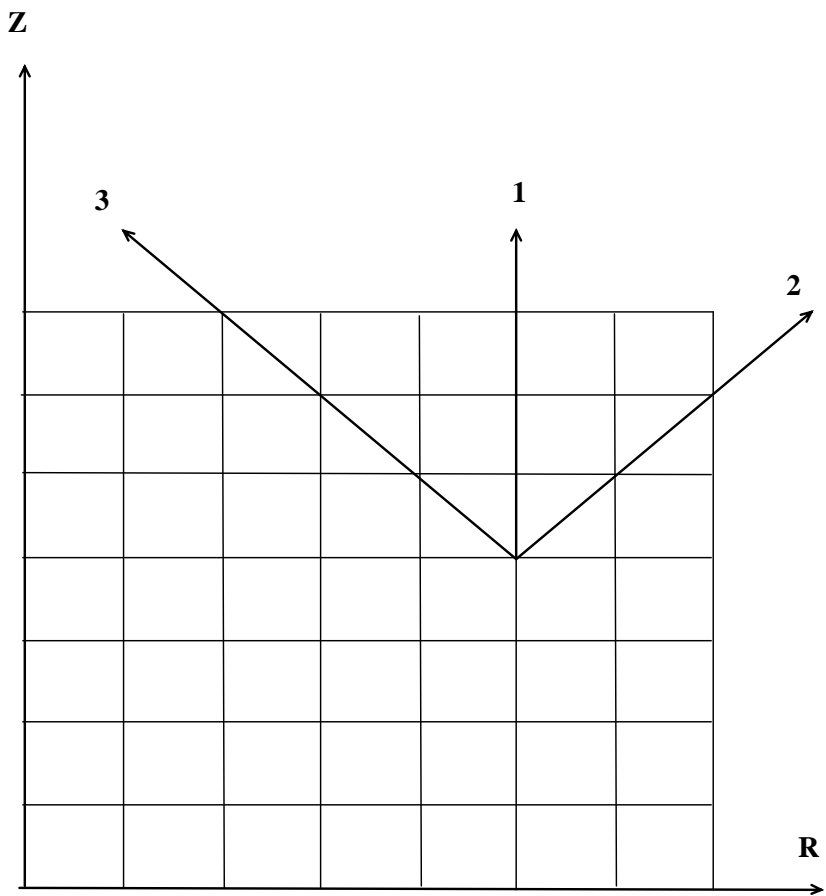,width=12cm,angle=-90}
    \caption{\label{nugrid} Cylindrical grid for neutrino opacities.}
\end{figure}

In order to evaluate the opacities along given directions we map the particle
properties density, temperature and electron fraction on an 
aequidistant, cylindrical grid with coordinates $(R,Z)$, where 
$R=\sqrt{x^2+y^2}$, see Figure \ref{nugrid}. 
The assumption of rotational symmetry around the
binary rotation axis is an excellent approximation since the main neutrino
emitting region is the hot, neutron star matter debris torus that forms around
the merged central object (see Fig. 14 in paper I). 
By evaluating the EOS at each grid point the 
matter properties (like the local composition) are known 
and we can therefore  assign a variable $\zeta_{\nu_i}$ (see eq.
(\ref{zeta})), containing compositional information, to each grid point. 
The neutrino grid does not have to be 
updated at every hydro time step. We chose to update it after a small 
fraction (1/8) of the neutron star dynamical time scale, $\tau_{dyn}= 
(G \bar{\rho})^{-1/2} \approx 2 \cdot 10^{-4}$ s, which is a tiny fraction
of the timescale on which typical disk properties change.
Once all the properties on the grid are known, the desired values at 
the SPH-particle positions are found by trilinear interpolation.
We use 400 points in radial direction and 300 points in 
positive Z-direction (symmetry with respect to the orbital plane is 
an excellent approximation for the systems under investigation).

The dominant sources of opacity are
\begin{itemize}
\item [(i)] neutrino nucleon scattering: 
   \begin{equation} \nu_i + \{n, p\} \rightarrow \nu_i + \{n, p\}
   \end{equation}
   with $\sigma_{\nu_i,nuc}= \frac{1}{4} \sigma_0 \left(\frac{E_{\nu_i}}
   {m_e c^2} \right)^2$ 
(Shapiro \& Teukolsky 1983) and
\item [(ii)] coherent neutrino nucleus scattering:
   	\begin{equation}
        \nu_i + A \rightarrow \nu_i + A \label{coh_scat_formula}
   	\end{equation}
	with $\sigma_{\nu_i,A}= \frac{1}{16} \sigma_0 
	\left(\frac{E_{\nu_i}}{m_e c^2}
 	\right)^2  A^2 (1-Z/A)^2$ 
        (Shapiro \& Teukolsky 1983; $\sin^2 \theta_W$ has 
	been approximated by 0.25). Here $A$ and $Z$ are the nucleon and 
	proton number of the average nucleus whose properties are stored in 
        our EOS-table. Due to the $A^2$-dependence of the cross section 
        this process will dominate as soon as a substantial fraction of heavy
	nuclei is present (remember that the nucleon numbers in these 
	nuclei reach values of up to $\sim$ 400 
        Shen et al. 1998).\\	
	Electron type neutrinos additionally undergo
\item [(iii)] neutrino absorption:
	\begin{eqnarray}
	\nu_e + n \rightarrow p + e^-\\
	\bar{\nu}_e + p \rightarrow n + e^+
	\end{eqnarray}
	with  $\sigma_{\nu_e,n}= \frac{1+3 \alpha^2}{4} \sigma_0 
	\left(\frac{E_{\nu}}{m_e c^2}
	 \right)^2 \langle 1-f_{e^-} \rangle$, where $\langle 1-f_{e^-} 
	\rangle \approx
	\left({\rm exp}(\eta_e - F_5(\eta_{\nu_e})/F_4(\eta_{\nu_e}))+1 \right)^{-1}$\\
	and  $\sigma_{\bar{\nu}_e,p}= \frac{1+3 \alpha^2}{4} \sigma_0 
	\left(\frac{E_{\bar{\nu}}}
	{m_e c^2} \right)^2 \langle 1-f_{e^+} \rangle$, 
         $\langle 1-f_{e^+} \rangle \approx
	\left({\rm exp}(-\eta_e - F_5(\eta_{\bar{\nu}_e})/F_4(\eta_{\bar{\nu}_e})) + 1)
	\right)^{-1}$.  
\end{itemize}

The local mean free path is given by (where for simplicity the spatial
dependence is suppressed)
\begin{equation} 
	\lambda_{\nu_i}(E)= \left(\sum_r n_r \sigma_r (E)\right)^{-1}\equiv
         (E^2 \zeta_{\nu_i})^{-1} \label{zeta},
\end{equation}
where the $n_r$ denote the target number densities, the index $r$ runs 
over the reactions given above with cross-sections $\sigma_r$ and $E$ is the
neutrino energy.
The dependence of the cross-sections on the squared neutrino energies has 
been separated out in the definition of $\zeta_{\nu_i}$. The optical depth,
$\tau$, along a specified direction is then given as 
\begin{equation} 
	\tau_{\nu_i}(E)= \int_{x1}^{x2} \frac{dx}{\lambda_{\nu_i}(E)}.
\end{equation}
The optical depths are evaluated along three directions from 
each grid point: in Z-direction ($\tau^1_{\nu_i}$), 
i.e. parallel to the rotational axis, along the outgoing diagonal 
($\tau^2_{\nu_i}$) and along the ingoing diagonal ($\tau^3_{\nu_i}$), see Fig.
\ref{nugrid}.
The finally used optical depth, ($\tau_{\nu_i}$), is the minimum of the  three,
$\tau_{\nu_i}= {\rm min} (\tau^1_{\nu_i},\tau^2_{\nu_i},\tau^3_{\nu_i}$).
The quantities that are actually stored for each grid point $j$ are
\begin{equation}
	\chi^d_{j,\nu_i}= \int_{d,j} \zeta_{\nu_i}(x) dx \label{chi},
\end{equation}
where $d$ denotes the direction and $\int_{d,j} dx$ is the integration from
grid point j along direction $d$. Note that the quantity $\chi$ is independent
of the neutrino energy and the (energy dependent) optical depth is given by
\begin{equation}
\tau_{\nu_i}(E)= E^2 {\rm min}_d (\chi_{j,\nu_i}^d) \equiv E^2 \chi_{j,\nu_i}.
\label{eq_tau_without_energy}
\end{equation}

The diffusion rate depends on the optical depth \( \tau _{\nu _{i}} \). We
base our estimates on a very simple, one-dimensional diffusion
model. Along one propagation direction we assume equal probabilities for
forward and backward scattering and impose strict flux conservation in a
stationary state situation. This leads to the following relationship between
the neutrino density \( J(E) \) and the neutrino number flux \( H(E) \),
\begin{equation}
\label{eq_diffusive_flux}
\frac{H_{\nu _{i}}(E)}{cJ_{\nu _{i}}(E)}=\frac{1}{2\tau _{\nu _{i}}(E)+1}.
\end{equation}
We can test this relationship against a complete numerical solution of
the diffusion equation in e.g. a supernova environment where all relevant
opacities are included and find agreement to about a factor of two. If the
thermodynamical conditions and the neutrino densities along the propagation
direction are set, relation
(\ref{eq_diffusive_flux}) defines a local neutrino number flux \( H_{\nu_{i}}(E) \)
which in general no longer obeys flux conservation in a stationary state
situation. Assuming that we still have a stationary state situation and that
the fluxes are locally well represented, we can use the balance of fluxes
across a infinitesimally thin layer perpendicular to the propagation direction
to obtain an 
estimate of the rate \( R_{\nu _{i}} \) of neutrinos produced in this layer.
Denoting the propagation direction with \( x \), we express the rate
in terms of the prevailing neutrino density and a diffusion time scale
\( T_{\nu _{i},x}^{dif} \) with
\begin{equation}
\label{eq_diffusion_rate}
R_{\nu _{i}}^{dif}(E)=\frac{\partial H_{\nu _{i}}(E)}{\partial x}=\frac{J_{\nu _{i}}(E)}{T_{\nu _{i}}^{dif}(E)}.
\end{equation}
The substitution of eq. (\ref{eq_diffusive_flux}) for \( H_{\nu _{i}} \)
leads to spatial derivatives of the neutrino density \( J_{\nu_{i}}(E) \) and the
optical depth \( \tau_{\nu_{i}} \). As the latter is given by the negative inverse
mean free path, \( -1/\lambda _{\nu _{i}}(E) \), eq.  (\ref{eq_diffusion_rate})
can be resolved for the diffusion time scale according to
\begin{equation}
T_{\nu _{i}}^{dif}(E)=\frac{2\tau _{\nu _{i}}(E)+1}{c}\left( \frac{\partial \ln J_{\nu _{i}}(E)}{\partial x}+\frac{2}{\left( 2\tau _{\nu _{i}}(E)+1\right) \lambda _{\nu _{i}}(E)}\right) ^{-1}.
\end{equation}
We rewrite this estimate with a distance parameter, \( \Delta x(E) \), to obtain
\begin{eqnarray}
T_{\nu _{i}}^{dif}(E) & = & \frac{\Delta x_{\nu _{i}}(E)}{c}\left( 2\tau _{\nu _{i}}(E)+1\right) ,\label{eq_diffusion_time_scale} \\
\Delta x_{\nu _{i}}(E) & = & \left( \frac{\partial \ln J_{\nu _{i}}(E)}{\partial x}+\frac{2}{\left( 2\tau _{\nu _{i}}(E)+1\right) \lambda _{\nu _{i}}(E)}\right) ^{-1}.\label{eq_diffusion_distance} 
\end{eqnarray}
The spatial derivative of the neutrino density in eq. 
(\ref{eq_diffusion_distance}) is quite inconvenient, one would prefer a diffusion time scale that does not
depend on neutrino densities. Moreover,  the derivative is likely to introduce noise
when evaluated in a three-dimensional numerical simulation. Hence, we 
neglect this term. In physical terms this means that we assume neutrino sources that keep
the neutrino density close to constant over a spatial interval where the mean free path
changes significantly. This might not always be justified and is subject to future
improvement. The expression for the distance parameter, however, greatly simplifies
to
\begin{equation}
\label{eq_diffusion_dist2}
\Delta x_{\nu _{i}}(E)=\left( \tau _{\nu _{i}}(E)+\frac{1}{2}\right) \lambda _{\nu _{i}}.
\end{equation}
Here we recall that Ruffert et al. (1996)
found the dependence
\begin{equation}
T_{\nu _{i}}^{dif}(E) = 3\frac{\Delta x_{\nu _{i}}(E)}{c}\tau _{\nu _{i}}(E)\label{Ttau}
\end{equation}
by calibration with a numerical neutrino transport scheme.
If we go back and use  \( \tau _{\nu _{i}}\sim 1 \) for the 
``last interaction region'' to simplify eq. (\ref{eq_diffusive_flux}) 
further by the approximation
\[
\frac{H_{\nu _{i}}(E)}{cJ_{\nu _{i}}(E)}=\frac{1}{3\tau _{\nu _{i}}(E)},
\]
we obtain eq. (\ref{Ttau}) by the same analysis used to derive eq. (\ref{eq_diffusion_time_scale}). However, the distance parameter is then given
by
\begin{equation}
\label{eq_diffusion_dist4}
\Delta x_{\nu _{i}}(E)=\tau _{\nu _{i}}(E)\lambda _{\nu _{i}}(E). 
\end{equation}
In our scheme  \( \Delta x \) defines the effective width of a layer drained by the
diffusive flux, i.e. provides the conversion between a net
emitted neutrino flux (number/s/cm\textasciicircum{}2)
and a production rate (number/s/cm\textasciicircum{}3).
We choose eqs. (\ref{Ttau}) and (\ref{eq_diffusion_dist4}) for our numerical
simulations because the linear dependence in \( \tau_{\nu_{i}} \) allows the
extraction of the energy dependence as in eq. (\ref{eq_tau_without_energy}).
Approximating the neutrino distribution function
in the high-density regime with a thermal equilibrium distribution we
apply the diffusion time scale and obtain the diffusion rates
\begin{equation}
	\langle R^{dif}_{\nu_i} \rangle = \int_0^\infty 
        \frac{\tilde{n}_{\nu_i}(E)}{T^{dif}_{\nu_i}(E)} dE =
	\frac{4 \pi c g_{\nu_i}}{(hc)^3} \frac{\zeta_{\nu_i}}{3 \chi^2_{\nu_i}} 
        T F_0(\eta_{\nu_i}) \label{Rdiff}
\end{equation}
\begin{equation}
	\langle Q^{dif}_{\nu_i} \rangle = \int_0^\infty 
        \frac{E \tilde{n}_{\nu_i}(E)}{T^{dif}_{\nu_i}(E)} dE =
	\frac{4 \pi c g_{\nu_i}}{(hc)^3} \frac{\zeta_{\nu_i}}{3 \chi^2_{\nu_i}} 
        T^2 F_1(\eta_{\nu_i}) \label{Qdiff}
\end{equation}
with
\begin{equation}
	\langle E^{dif}_{\nu_i} \rangle = 
        \frac{\langle Q^{dif}_{\nu_i} \rangle}{\langle R^{dif}_{\nu_i} \rangle}
	= T \frac{F_1(\eta_{\nu_i})}{F_0(\eta_{\nu_i})}.
\end{equation}
Here $\tilde{n}_{\nu_i}(E)$ is related to the number density 
by $n_{\nu_i}= \int_0^\infty \tilde{n}_{\nu_i}(E) dE$.
The statistical weights $g_{\nu_i}$ are 1 for $\nu_e$ and $\bar{\nu}_e$
and 4 for $\nu_x$. After the second equals sign in eqs. (\ref{Rdiff}) and 
(\ref{Qdiff}) we have inserted the explicit estimate (\ref{Ttau}) for 
the diffusion time scale and (\ref{eq_diffusion_dist4}) for the distance
parameter. 
Note that this leakage prescription is  {\em not} based on the use of 
mean neutrino energies, eq. (\ref{Eeff}) is exclusively used for informative 
purposes. Our scheme accounts for the energy dependence of the 
neutrino opacities by integrating over the neutrino distribution.

\end{appendix}



\begin{thebibliography}{}
\bibitem[\protect\citeauthoryear{Abramovici, {Althouse}, {Drever}, {Gursel},
  {Kawamura}, {Raab}, {Shoemaker}, {Sievers}, {Spero} \& {Thorne}}{Abramovici
  et~al.}{1992}]{abramovici92}
Abramovici A.,  {Althouse} W.~E.,  {Drever} R. W.~P.,  {Gursel} Y.,  {Kawamura}
  S.,  {Raab} F.~J.,  {Shoemaker} D.,  {Sievers} L.,  {Spero} R.~E.,
  {Thorne} K.~S.,  1992, Science, 256, 325



\bibitem{Arnett_67}Arnett, W.~D. 1967, Canadian J. of Phys., 215, 1621
\bibitem{Arnett_77}Arnett, W.~D. 1977, ApJ, 218, 815



\bibitem[\protect\citeauthoryear{{Ayal}, {Piran}, {Oechslin}, {Davies} \&
  {Rosswog}}{{Ayal} et~al.}{2001}]{ayal01}
{Ayal} S.,  {Piran} T.,  {Oechslin} R.,  {Davies} M.~B.,    {Rosswog} S.,
  2001, ApJ, 550, 846


\bibitem{balsara95}
Balsara D.,  1995, J. Comput. Phys., 121, 357


\bibitem{Baron_Cooperstein_Kahana_85b}Baron E., Cooperstein J., \& Kahana, S. 1985, Phys. Rev. Lett., 55,
126


\bibitem[\protect\citeauthoryear{Baumgarte, Cook, Scheel, Shapiro \&
  Teukolsky}{Baumgarte et~al.}{1997}]{baumgarte97}
Baumgarte T.,  Cook G.,  Scheel M.,  Shapiro S.,    Teukolsky S.,  1997, Phys.
  Rev. Lett., 79, 1182

\bibitem{benz90a}
Benz W.,  1990, in Buchler J.,  ed.,  Numerical Modeling of Stellar
  Pulsations.
Kluwer Academic Publishers, Dordrecht, p.~269

\bibitem{benz90b}
Benz W.,  Bowers R., Cameron A., Press W.,  1990, ApJ, 348, 647

\bibitem{Bethe_Wilson_85}Bethe, H.~A. \& Wilson, J.~R. 1985, ApJ, 295, 14


\bibitem{bildsten92}
Bildsten, L. \& Cutler, C., 1992, ApJ, 400, 175



\bibitem{Bowers_Wilson_82}
Bowers, R.~L. \& Wilson, J.~R. 1982, ApJS, 50, 115

\bibitem[\protect\citeauthoryear{Bradaschia et~al.}{Bradaschia et~al.}{1990}]{bradaschia90}
Bradaschia C. et~al.,  1990, Nucl.Instrum. Methods Phys. Res. A, 289, 518

\bibitem{bruenn85}
Bruenn, S.W., 1985, ApJS, 58, 771

\bibitem{Bruenn_DeNisco_Mezzacappa_01}Bruenn, S.~W., DeNisco, K.~R., \& Mezzacappa, A. 2001, ApJ, 560,
326
\bibitem{Burrows_Hayes_Fryxell_95}Burrows, A., Hayes, J., \& Fryxell, B.~A. 1995, ApJ, 450, 830
\bibitem{Burrows_et_al_00}Burrows, A., Young, T., Pinto, Ph., Eastman, R. \& Thompson, T.~A.
2000, ApJ, 539, 865
\bibitem{Burrows02}Burrows, A., Thompson, T.~A., Pinto, Ph. 2002,  
astro-ph/0211194
\bibitem{Colgate_White_66}Colgate, S.~A. \& White, R.~H. 1966, ApJ, \textbf{}143, 626
\bibitem{cooperstein86}
Cooperstein, J., van den Horn, L. J., Baron, E. A., 1986, ApJ, 309, 653


\bibitem[\protect\citeauthoryear{Danzmann}{Danzmann}{1997}]{danzmann97}
Danzmann K.,  1997, in of Sciences T. N. Y.~A.,  ed., Proceedings of the
  17$^{th}$ Texas Symposium on relativistic astrophysics and cosmology New York


\bibitem{eichler89}
Eichler D., Livio M.,  Piran T. \& Schramm D.N., 1989,
Nature 340, 126 

\bibitem[\protect\citeauthoryear{Faber \& Rasio}{Faber \&
  Rasio}{2000}]{faber00}
Faber J.,  Rasio F.,  2000, Phys.Rev. D62, p. 064012

\bibitem[\protect\citeauthoryear{Faber, Rasio \& Manor}{Faber
  et~al.}{2001}]{faber01}
Faber J.,  Rasio F.,    Manor J.,  2001, Phys.Rev. D63, p. 044012


\bibitem{Fryer_Warren_02}Fryer, C.~F. \& Warren, M.~S. 2002, ApJL, 574, 65


\bibitem[\protect\citeauthoryear{Freiburghaus, Rosswog \&
  Thielemann}{Freiburghaus et~al.}{1999}]{freiburghaus99b}
Freiburghaus C.,  Rosswog S.,    Thielemann F.-K.,  1999, ApJ, 525, L121



\bibitem{haft94}
Haft, M.,  Raffelt, G. \& Weiss, A., 1994, ApJ, 425

\bibitem{Herant_Benz_Colgate_92}Herant, M., Benz, W., \& Colgate, S.~A. 1992, ApJ, 395, 642
\bibitem{Herant_et_al_94}Herant M., Benz W., Hix R.~W., Fryer C.~L., \& Colgate, S.~A. 1994,
ApJ, 435, 339

\bibitem{itoh96}
Itoh, N., Nishikawa, A. \& Kohyama, Y., 1996, ApJ, 470, 1015

\bibitem{Janka_Mueller_96}Janka, H.-T. \& M\"{u}ller, E. 1996, A\&A, 306, 167
\bibitem{Janka_Kifonidis_Rampp_01}Janka, H.-T., Kifonidis, K., \& Rampp, M. 2001, in Proc. Workshop
on Physics of Neutron Star Interiors, ed. D. Blaschke, N. Glendenning,
\& A. Sedrakian, Lecture Notes in Physics (Germany: Springer), 333


\bibitem{kluzniak98}
Kluzniak, W. \& Ruderman, 1998, ApJ, 508, L113

\bibitem{kochanek92}
Kochanek, C.S., 1992, ApJ, 398, 234


\bibitem[\protect\citeauthoryear{Kuroda et~al.}{Kuroda et~al.}{1997}]{kuroda97}
{Kuroda} K. et~al., 1997, in Gravitational Wave
  Detection, Proceedings of the TAMA International Workshop on Gravitational
  Wave Detection held at National Women's Education Centre, Saitama, Japan, on
  12-14 November, 1996. Edited by K. Tsubono, M.-K. Fujimoto, and K. Kuroda.
  Frontiers Science Series No. 20. Universal Academy Press, Inc., 1997., p.309
  Japanese gravitational wave observatory (jgwo).
pp~309+

\bibitem{lai94}
Lai, D. 1994, MNRAS, 270, 611

\bibitem[\protect\citeauthoryear{Lattimer \& Schramm}{Lattimer \&
  Schramm}{1974}]{lattimer74}
Lattimer J.~M.,  Schramm D.~N.,  1974, ApJ, (Letters), 192, L145

\bibitem[\protect\citeauthoryear{Lattimer \& Schramm}{Lattimer \&
  Schramm}{1976}]{lattimer76}
Lattimer J.~M.,  Schramm D.~N.,  1976, ApJ, 210, 549


Inc.)
\bibitem{Liebendoerfer_et_al_01}Liebend\"{o}rfer, M., Mezzacappa, A., Thielemann, F.-K., Messer,
O.~E.~B., Hix, W.~R., \& Bruenn, S.~W. 2001, Phys. Rev., D63,
103004
\bibitem{Liebendoerfer_et_al_02b}Liebend\"{o}rfer, M., Messer, O.~.E.~B., Mezzacappa, A., Bruenn, S.W., Cardall, C.Y., Thielemann, F.K., 2002, submitted to ApJS, 
astro-ph/0207036

\bibitem{lyford02}Lyford, N.D., Baumgarte, T.W., Shapiro, S.L., 2002, to
appear in ApJ, gr-qc/0210012







\bibitem{Mezzacappa_Bruenn_93a}Mezzacappa, A. \& Bruenn, S.~W. 1993, ApJ 405, 669
\bibitem{Mezzacappa_et_al_98b}Mezzacappa, A., Calder, A.~C., Bruenn, S.~W., Blondin, J.~M., Guidry,
M.~W., Strayer, M.~R., \& Umar, A.~S. 1998, ApJ, 495, 911
\bibitem{Mezzacappa_Messer_98}Mezzacappa, A. \& Messer, O.~E.~B. 1999, JCAM, 109, 281
\bibitem{Mezzacappa_et_al_01}Mezzacappa, A., Liebend\"{o}rfer, M., Messer, O.~E.~B., Hix, W.~R.,
Thielemann, F.-K., \& Bruenn, S.~W. 2001, PRL, 86, 1935

\bibitem{monaghan92}
Monaghan J.J.,  1992, Ann. Rev. Astron. Astrophys., 30, 543

\bibitem{monaghan97}
Morris J.,  Monaghan J.J.,  1997, J. Comp. Phys., 136, 41
\bibitem{Myra_et_al_87}Myra, E.~S., Bludman, S.~A., Hoffman, Y., Lichtenstadt, I., Sack,
N., \& Van Riper, K.~A. 1987, ApJ, 318, 744

\bibitem[Narayan et al.(1992)]{narayan92} Narayan R., Paczy\'{n}ski
B., Piran T., 1992, ApJ, 395, L83

\bibitem[\protect\citeauthoryear{Oechslin, Rosswog \& Thielemann}{Oechslin
  et~al.}{2001}]{oechslin01}
Oechslin R.,  Rosswog S.,    Thielemann F.-K.,  2002, Phys. Rev. D, 65, 103005


\bibitem[\protect\citeauthoryear{Oohara \& Nakamura}{Oohara \&
  Nakamura}{1997}]{oohara97}
Oohara K.,  Nakamura T.,  1997, in Relativistic Gravitation and Gravitational
  Radiation.
Cambridge University Press, Cambridge

\bibitem[Paczy\'{n}ski(1986)]{paczynski86} Paczy\'{n}ski B., 1986, 
ApJ, 308, L43

\bibitem{Pons_et_al_99}Pons, J.~A., Reddy, S., Prakash, M., Lattimer, J.~M., \& Miralles,
J.~A. 1999, ApJ, 513, 780



\bibitem{Rampp_Janka_00}Rampp, M. \& Janka, H.~T. 2000, ApJL, 539, L33

\bibitem[\protect\citeauthoryear{Rosswog, Liebend\"orfer, Thielemann, Davies,
  Benz \& Piran}{Rosswog et~al.}{1999}]{rosswog99}
Rosswog S.,  Liebend\"orfer M.,  Thielemann F.-K.,  Davies M.~B.,  Benz W.,
  Piran T.,  1999, A \&\ A, 341, 499

\bibitem{rosswog00}
Rosswog S.,  Davies M.~B.,  Thielemann F.-K.,    Piran T.,  2000, A \&\ A, 360,
  171

\bibitem[Rosswog \& Davies 2002]{rosswog02a}
Rosswog S.,  Davies M.~B.,  2002, MNRAS, 334, 481

\bibitem[Rosswog \& Ramirez-Ruiz E. 2002]{rosswog02b} Rosswog S., Ramirez-Ruiz E.,
2002, MNRAS, 336, L7

\bibitem[\protect\citeauthoryear{Ruffert, Janka \& Sch\"afer}{Ruffert
  et~al.}{1996}]{ruffert96}
Ruffert M.,  Janka H.,    Sch\"afer G.,  1996, A \&\ A, 311, 532

\bibitem[Ruffert et al. 1997a]{ruffert97a}
Ruffert M., Janka H.-T., Sch\"afer G., 1997, A \& A, 311, 532 



\bibitem[Ruffert \& Janka 2001]{ruffert01}
Ruffert M., Janka H.-T., 2001, A \&\ A, 380, 544 


\bibitem{Schwartz_67}Schwartz, R.~A. 1967, Ann. Phys., 43, 42

\bibitem{shapiro83}
Shapiro,S. \& Teukolsky, S.A., 1983, 
Black holes, White Dwarfs and Neutron Stars, (New York) Whiley \& Sons


\bibitem{shen98a}
Shen H.,  Toki H.,  Oyamatsu K.,    Sumiyoshi K.,  1998a, Nuclear Physics, A
  637, 435

\bibitem{shen98b}
Shen H.,  Toki H.,  Oyamatsu K.,    Sumiyoshi K.,  1998b, Prog. Theor. Phys.,
100, 1013

\bibitem[\protect\citeauthoryear{Shibata}{Shibata}{1999}]{shibata99}
Shibata M.,  1999, Phys. Rev. D, 60, 104052

\bibitem[\protect\citeauthoryear{Shibata \& Uryu}{Shibata \&
  Uryu}{2000}]{shibata00}
Shibata M.,  Uryu K.,  2000, Phys. Rev. D, 61, 064001

\bibitem[\protect\citeauthoryear{Shibata \& Uryu}{Shibata \&
  Uryu}{2002}]{shibata02}
Shibata M.,  Uryu K.,  2002, Prog. Theor. Phys., 2002, 107, 265

356, 559

\bibitem[\protect\citeauthoryear{Symbalisty \& Schramm}{Symbalisty \&
  Schramm}{1982}]{symbalisty82}
Symbalisty E. M.~D.,  Schramm D.~N.,  1982, Astrophys. Lett., 22, 143

\bibitem{takahashi78}
Takahashi, K., El Eid, M.F. \& Hillebrandt, W., 1978, A\&A, 67, 185

\bibitem{taylor94}Taylor, J.H., 1994, Rev. Mod. Phys., 66, 711 

\bibitem{thompson00}
Thompson, T.A., Burrows, A. \& Horvath, J.E., 2000, Phys. Rev. C, 62, 03580

\bibitem{thompson93}
Thompson, C. and Duncan, R.C., 1993, ApJ, 408, 194

\bibitem{thompson94}
Thompson, C., 1994, MNRAS, 270, 480

\bibitem{thompson01}
Thompson, T.A., Burrows,A., Meyer, B.S., 2001, ApJ, 562, 887 

\bibitem{thorsett99}
Thorsett, S.E. \& Chakrabarti, D., 1999, ApJ, 512, 288

\bibitem{tubbs75}
Tubbs, D.L. \& Schramm, D.N., 1975, ApJ, 201, 467
\bibitem{VanRiper_79}Van Riper, K.~A. 1979, ApJ, 232, 558
\bibitem{VanRiper_Lattimer_81}Van Riper, K.~A., \& Lattimer, J.~M. 1981, ApJ, 249, 270
\bibitem{Wilson_71}Wilson, J.~R. 1971, ApJ, 163, 209
\bibitem{Wilson_85}Wilson, J.~R. 1985, in Numerical Astrophysics, ed. by Centrella,
J.~M., LeBlanc, J.~M., \& Bowers, R.~L. (Boston: Jones and Bartlett)


\bibitem[\protect\citeauthoryear{Wilson et al.}{Wilson et al.}{1996}]{wilson96}
Wilson J.~R.,  Mathews G., \& Marronetti, P.  1996, Phys. Rev. D, 54, 1317


\end{thebibliography}



\clearpage

\clearpage

\clearpage

\clearpage

\clearpage

\clearpage

\clearpage

\clearpage

\clearpage

\clearpage

\clearpage

\clearpage

\end{document}